\documentclass[twocolumn,trackchanges]{aastex631}

\newcommand{\degree}{$^{\circ}$}

\usepackage{natbib}
\usepackage{lmodern}
\usepackage{amsmath}
\usepackage{float}
\usepackage{svg}
\usepackage{comment}
\begin{document}

\title{TeV Analysis of a Source Rich Region with HAWC Observatory: Is HESS J1809-193 a Potential Hadronic PeVatron?}
\author{A.~Albert}
\affiliation{Physics Division, Los Alamos National Laboratory, Los Alamos, NM, USA }

\author{R.~Alfaro}
\affiliation{Instituto de F'{i}sica, Universidad Nacional Autónoma de México, Ciudad de Mexico, Mexico }

\author{C.~Alvarez}
\affiliation{Universidad Autónoma de Chiapas, Tuxtla Gutiérrez, Chiapas, México}

\author{J.C.~Arteaga-Velázquez}
\affiliation{Universidad Michoacana de San Nicolás de Hidalgo, Morelia, Mexico }

\author{D.~Avila Rojas}
\affiliation{Instituto de F'{i}sica, Universidad Nacional Autónoma de México, Ciudad de Mexico, Mexico }

\author{R.~Babu}
\affiliation{Department of Physics, Michigan Technological University, Houghton, MI, USA }
\affiliation{Department of Physics and Astronomy, Michigan State University, East Lansing, MI, USA}

\author{E.~Belmont-Moreno}
\affiliation{Instituto de F'{i}sica, Universidad Nacional Autónoma de México, Ciudad de Mexico, Mexico }

\author{A.~Bernal}
\affiliation{Instituto de Astronom'{i}a, Universidad Nacional Autónoma de México, Ciudad de Mexico, Mexico }

\author{M.~Breuhaus}
\affiliation{Max-Planck Institute for Nuclear Physics, 69117 Heidelberg, Germany}

\author{K.S.~Caballero-Mora}
\affiliation{Universidad Autónoma de Chiapas, Tuxtla Gutiérrez, Chiapas, México}

\author{T.~Capistrán}
\affiliation{Instituto de Astronom'{i}a, Universidad Nacional Autónoma de México, Ciudad de Mexico, Mexico }

\author{A.~Carramiñana}
\affiliation{Instituto Nacional de Astrof'{i}sica, Óptica y Electrónica, Puebla, Mexico }

\author{S.~Casanova}
\affiliation{Instytut Fizyki Jadrowej im Henryka Niewodniczanskiego Polskiej Akademii Nauk, IFJ-PAN, Krakow, Poland }

\author{J.~Cotzomi}
\affiliation{Facultad de Ciencias F'{i}sico Matemáticas, Benemérita Universidad Autónoma de Puebla, Puebla, Mexico }

\author{E.~De la Fuente}
\affiliation{Departamento de F'{i}sica, Centro Universitario de Ciencias Exactase Ingenierias, Universidad de Guadalajara, Guadalajara, Mexico }

\author{D.~Depaoli}
\affiliation{Max-Planck Institute for Nuclear Physics, 69117 Heidelberg, Germany}

\author{N.~Di Lalla}
\affiliation{Department of Physics, Stanford University: Stanford, CA 94305–4060, USA}

\author{R.~Diaz Hernandez}
\affiliation{Instituto Nacional de Astrof'{i}sica, Óptica y Electrónica, Puebla, Mexico }

\author{B.L.~Dingus}
\affiliation{Physics Division, Los Alamos National Laboratory, Los Alamos, NM, USA }

\author{M.A.~DuVernois}
\affiliation{Department of Physics, University of Wisconsin-Madison, Madison, WI, USA }

\author{C.~Espinoza}
\affiliation{Instituto de F'{i}sica, Universidad Nacional Autónoma de México, Ciudad de Mexico, Mexico }

\author{K.L.~Fan}
\affiliation{Department of Physics, University of Maryland, College Park, MD, USA }

\author{K.~Fang}
\affiliation{Department of Physics, University of Wisconsin-Madison, Madison, WI, USA }

\author{B.~Fick}
\affiliation{Department of Physics, Michigan Technological University, Houghton, MI, USA }

\author{N.~Fraija}
\affiliation{Instituto de Astronom'{i}a, Universidad Nacional Autónoma de México, Ciudad de Mexico, Mexico }

\author{J.A.~García-González}
\affiliation{Tecnologico de Monterrey, Escuela de Ingenier\'{i}a y Ciencias, Ave. Eugenio Garza Sada 2501, Monterrey, N.L., Mexico, 64849}

\author{F.~Garfias}
\affiliation{Instituto de Astronom'{i}a, Universidad Nacional Autónoma de México, Ciudad de Mexico, Mexico }

\author{A.~Gonzalez Muñoz}
\affiliation{Instituto de F'{i}sica, Universidad Nacional Autónoma de México, Ciudad de Mexico, Mexico }

\author{M.M.~González}
\affiliation{Instituto de Astronom'{i}a, Universidad Nacional Autónoma de México, Ciudad de Mexico, Mexico }

\author{J.A.~Goodman}
\affiliation{Department of Physics, University of Maryland, College Park, MD, USA }

\author{S.~Groetsch}
\affiliation{Department of Physics, Michigan Technological University, Houghton, MI, USA }

\author{J.P.~Harding}
\affiliation{Physics Division, Los Alamos National Laboratory, Los Alamos, NM, USA }

\author{S.~Hernández-Cadena}
\affiliation{Instituto de F'{i}sica, Universidad Nacional Autónoma de México, Ciudad de Mexico, Mexico }

\author{I.~Herzog}
\affiliation{Department of Physics and Astronomy, Michigan State University, East Lansing, MI, USA }

\author{D.~Huang}
\affiliation{Department of Physics, University of Maryland, College Park, MD, USA }

\author{F.~Hueyotl-Zahuantitla}
\affiliation{Universidad Autónoma de Chiapas, Tuxtla Gutiérrez, Chiapas, México}

\author{P.~Hüntemeyer}
\affiliation{Department of Physics, Michigan Technological University, Houghton, MI, USA }

\author{A.~Iriarte}
\affiliation{Instituto de Astronom'{i}a, Universidad Nacional Autónoma de México, Ciudad de Mexico, Mexico }

\author{V.~Joshi}
\affiliation{Erlangen Centre for Astroparticle Physics, Friedrich-Alexander-Universit\"at Erlangen-N\"urnberg, Erlangen, Germany}

\author{S.~Kaufmann}
\affiliation{Universidad Politecnica de Pachuca, Pachuca, Hgo, Mexico }

\author{A.~Lara}
\affiliation{Instituto de Geof'{i}sica, Universidad Nacional Autónoma de México, Ciudad de Mexico, Mexico }

\author{J.~Lee}
\affiliation{University of Seoul, Seoul, Rep. of Korea}

\author{H.~León Vargas}
\affiliation{Instituto de F'{i}sica, Universidad Nacional Autónoma de México, Ciudad de Mexico, Mexico }

\author{A.L.~Longinotti}
\affiliation{Instituto de Astronom'{i}a, Universidad Nacional Autónoma de México, Ciudad de Mexico, Mexico }

\author{G.~Luis-Raya}
\affiliation{Universidad Politecnica de Pachuca, Pachuca, Hgo, Mexico }

\author{K.~Malone}
\affiliation{Physics Division, Los Alamos National Laboratory, Los Alamos, NM, USA }

\author{J.~Martínez-Castro}
\affiliation{Centro de Investigaci'on en Computaci'on, Instituto Polit'ecnico Nacional, M'exico City, M'exico.}

\author{J.A.~Matthews}
\affiliation{Dept of Physics and Astronomy, University of New Mexico, Albuquerque, NM, USA }

\author{P.~Miranda-Romagnoli}
\affiliation{Universidad Autónoma del Estado de Hidalgo, Pachuca, Mexico }

\author{J.A.~Montes}
\affiliation{Instituto de Astronom'{i}a, Universidad Nacional Autónoma de México, Ciudad de Mexico, Mexico }

\author{J.A.~Morales-Soto}
\affiliation{Universidad Michoacana de San Nicolás de Hidalgo, Morelia, Mexico }

\author{E.~Moreno}
\affiliation{Facultad de Ciencias F'{i}sico Matemáticas, Benemérita Universidad Autónoma de Puebla, Puebla, Mexico }

\author{M.~Mostafá}
\affiliation{Department of Physics, Temple University, Philadelphia, Pennsylvania, USA}

\author{L.~Nellen}
\affiliation{Instituto de Ciencias Nucleares, Universidad Nacional Autónoma de Mexico, Ciudad de Mexico, Mexico }

\author{M.~Newbold}
\affiliation{Department of Physics and Astronomy, University of Utah, Salt Lake City, UT, USA }

\author{M.U.~Nisa}
\affiliation{Department of Physics and Astronomy, Michigan State University, East Lansing, MI, USA }

\author{R.~Noriega-Papaqui}
\affiliation{Universidad Autónoma del Estado de Hidalgo, Pachuca, Mexico }

\author{M.~Osorio}
\affiliation{Instituto de Astronom'{i}a, Universidad Nacional Autónoma de México, Ciudad de Mexico, Mexico}

\author{Y.~Pérez Araujo}
\affiliation{Instituto de F'{i}sica, Universidad Nacional Autónoma de México, Ciudad de Mexico, Mexico }

\author{E.G.~Pérez-Pérez}
\affiliation{Universidad Politecnica de Pachuca, Pachuca, Hgo, Mexico }

\author{C.D.~Rho}
\affiliation{Department of Physics, Sungkyunkwan University, Suwon 16419, South Korea}

\author{D.~Rosa-González}
\affiliation{Instituto Nacional de Astrof'{i}sica, Óptica y Electrónica, Puebla, Mexico }

\author{E.~Ruiz-Velasco}
\affiliation{Max-Planck Institute for Nuclear Physics, 69117 Heidelberg, Germany}

\author{H.~Salazar}
\affiliation{Facultad de Ciencias F'{i}sico Matemáticas, Benemérita Universidad Autónoma de Puebla, Puebla, Mexico }

\author{A.~Sandoval}
\affiliation{Instituto de F'{i}sica, Universidad Nacional Autónoma de México, Ciudad de Mexico, Mexico }

\author{M.~Schneider}
\affiliation{Department of Physics, University of Maryland, College Park, MD, USA }

\author{J.~Serna-Franco}
\affiliation{Instituto de F'{i}sica, Universidad Nacional Autónoma de México, Ciudad de Mexico, Mexico }

\author{A.J.~Smith}
\affiliation{Department of Physics, University of Maryland, College Park, MD, USA }

\author{Y.~Son}
\affiliation{University of Seoul, Seoul, Rep. of Korea}

\author{R.W.~Springer}
\affiliation{Department of Physics and Astronomy, University of Utah, Salt Lake City, UT, USA }

\author{O.~Tibolla}
\affiliation{Universidad Politecnica de Pachuca, Pachuca, Hgo, Mexico }

\author{K.~Tollefson}
\affiliation{Department of Physics and Astronomy, Michigan State University, East Lansing, MI, USA }

\author{I.~Torres}
\affiliation{Instituto Nacional de Astrof'{i}sica, Óptica y Electrónica, Puebla, Mexico }

\author{R.~Torres-Escobedo}
\affiliation{Tsung-Dao Lee Institute \& School of Physics and Astronomy, Shanghai Jiao Tong University, Shanghai, China}

\author{R.~Turner}
\affiliation{Department of Physics, Michigan Technological University, Houghton, MI, USA }

\author{F.~Ureña-Mena}
\affiliation{Instituto Nacional de Astrof'{i}sica, Óptica y Electrónica, Puebla, Mexico }

\author{E.~Varela}
\affiliation{Facultad de Ciencias F'{i}sico Matemáticas, Benemérita Universidad Autónoma de Puebla, Puebla, Mexico }

\author{X.~Wang}
\affiliation{Department of Physics, Michigan Technological University, Houghton, MI, USA }

\author{I.J.~Watson}
\affiliation{University of Seoul, Seoul, Rep. of Korea}

\author{E.~Willox}
\affiliation{Department of Physics, University of Maryland, College Park, MD, USA }

\author{S.~Yun-Cárcamo}
\affiliation{Department of Physics, University of Maryland, College Park, MD, USA }

\author{H.~Zhou}
\affiliation{Tsung-Dao Lee Institute \& School of Physics and Astronomy, Shanghai Jiao Tong University, Shanghai, China}
\collaboration{1000}{(THE HAWC COLLABORATION)}

\correspondingauthor{R. Babu}
\email{rbabu@mtu.edu}

\begin{abstract}

HESS J1809-193 is an unidentified TeV source, first detected by the High Energy Stereoscopic System (H.E.S.S.) Collaboration. The emission originates in a source-rich region that includes several Supernova Remnants (SNR) and Pulsars (PSR) including SNR G11.1+0.1, SNR G11.0-0.0, and the young radio pulsar J1809-1917. Originally classified as a pulsar wind nebula (PWN) candidate, recent studies show the peak of the TeV region overlapping with a system of molecular clouds. This resulted in the revision of the original leptonic scenario to look for alternate hadronic scenarios. Marked as a potential PeVatron candidate, this region has been studied extensively by H.E.S.S. due to its emission extending up-to several tens of TeV. In this work, we use 2398 days of data from the High Altitude Water Cherenkov (HAWC) observatory to carry out a systematic source search for the HESS J1809-193 region. We were able to resolve emission detected as an extended component (modelled as a Symmetric Gaussian with a 1$\sigma$ radius of 0.21$^{\circ}$) with no clear cutoff at high energies and emitting photons up-to 210 TeV. We model the multi-wavelength observations for the region HESS J1809-193 using a time-dependent leptonic model and a lepto-hadronic model. Our model indicates that both scenarios could explain the observed data within the region of HESS J1809-193.

\end{abstract}

\section{Introduction}
\subsection{Previous TeV measurements}
HESS J1809-193 was originally discovered by the High Energy Stereoscopic System (H.E.S.S.) detector in 2007\citep{Originpaper} as a part of a systematic search for very-high-energy emission from energetic pulsars in the Galactic plane in the very-high-energy range (up to 30TeV). H.E.S.S. originally reported that it is an extended source with a fairly hard spectral index ($\sim$2.2) that could be possibly associated with a Pulsar Wind Nebula (PWN) powered by the pulsar PSR J1809-1917. 
\par
HESS J1809-193 was detected by the High Altitude Water Cherenkov Detector (HAWC) Observatory in the HAWC second source catalog on TeV sources \citep{2HWC} as 2HWC J1809-190 and also in the third catalog of TeV sources \citep{3HWC} with the source name as 3HWC J1809-190. The emission of 3HWC J1809-190 is centered at (R.A, Decl.) of (272.46$^\circ$, -19.04$^\circ$) with a 16$\sigma$ significance(pre-trials) using 1523 days of HAWC data.
\par
The H.E.S.S. Collaboration, in 2023, updated their observations with a total of 93.2 hours of observation time above 0.27 TeV \citep{new1809hesspaper}. They were able to resolve the emission into two components: an extended Component A, and a compact Component B. The extended Component A was modelled as an elongated Gaussian with a 1$\sigma$ semi-major/semi-minor axis of $\sim$ 0.62$^\circ$ / $\sim$ 0.35$^\circ$, that shows a spectral cutoff at $\sim$ 12 TeV. Component B is modelled as a symmetric Gaussian with a 1$\sigma$ radius of $\sim$ 0.1$^\circ$ modelled with a hard spectrum and shows no spectral cutoff. The location of Component B is closer to PSR J1809-1917. \citet{new1809hesspaper} modelled the region using a time-dependent leptonic scenario based on three generations of electrons: a halo of `relic' electrons associated with the extended Component A, `medium-age' electrons associated with the compact Component B and the `young' electrons associated with an X-ray nebula\citep{xraysuzaku}. 
\par
Fermi-LAT observations of the region, performed by \citet{araya} and \citet{new1809hesspaper} using 3FGL and 4FGL data releases respectively, list two sources in the region. As noted in \citet{new1809hesspaper}, the two sources in the region are:(i) 4FGL J1810.3-1925e, which is modelled as an extended source(Gaussian morphology) with a LogParabola spectral curvature model, (ii)  4FGL J1811.5-1925 is modelled as a point source and a power-law spectral model.
4FGL J1810.3-1925e's best-fit position is closer to PSR J1809-1917 and the two H.E.S.S. components, indicating an association with the emission observed by H.E.S.S. The extension of the Fermi source ($\sigma\sim0.3^\circ$) is also comparable to the extension of the extended H.E.S.S. Component A ($\sigma\sim0.6^\circ$). 
4FGL J1811.5-1925 is positionally coincident with PSR J1811-1925, which indicates the association of the source with the pulsar. Therefore the emission from this region is not considered to contribute towards the bulk emission from HESS J1809-193.
\par
HESS J1809-193 is also reported by the Large High Altitude Air Shower Observatory (LHAASO) Collaboration in 2023 as an ultra high-energy source 1LHAASO J1809-1918u \citep{lhaasocat}. LHAASO WCDA array detected the region as an extended gaussian with a 1$\sigma$ radius of $\sim$ 0.35$^\circ$, with a spectral index of 2.24 between 1-25 TeV. The KM2A array detects the region as a point source with a 1$\sigma$ radius of $\sim$ 0.22$^\circ$ as an upper limit, with a soft spectral index of 3.51, at energies above 25 TeV. The source was detected by KM2A with a  significance of 9.4 $\sigma$ at energies greater than 100 TeV, it is marked as an ultra-high-energy source.
\par
\begin{figure}[ht!]
\fig{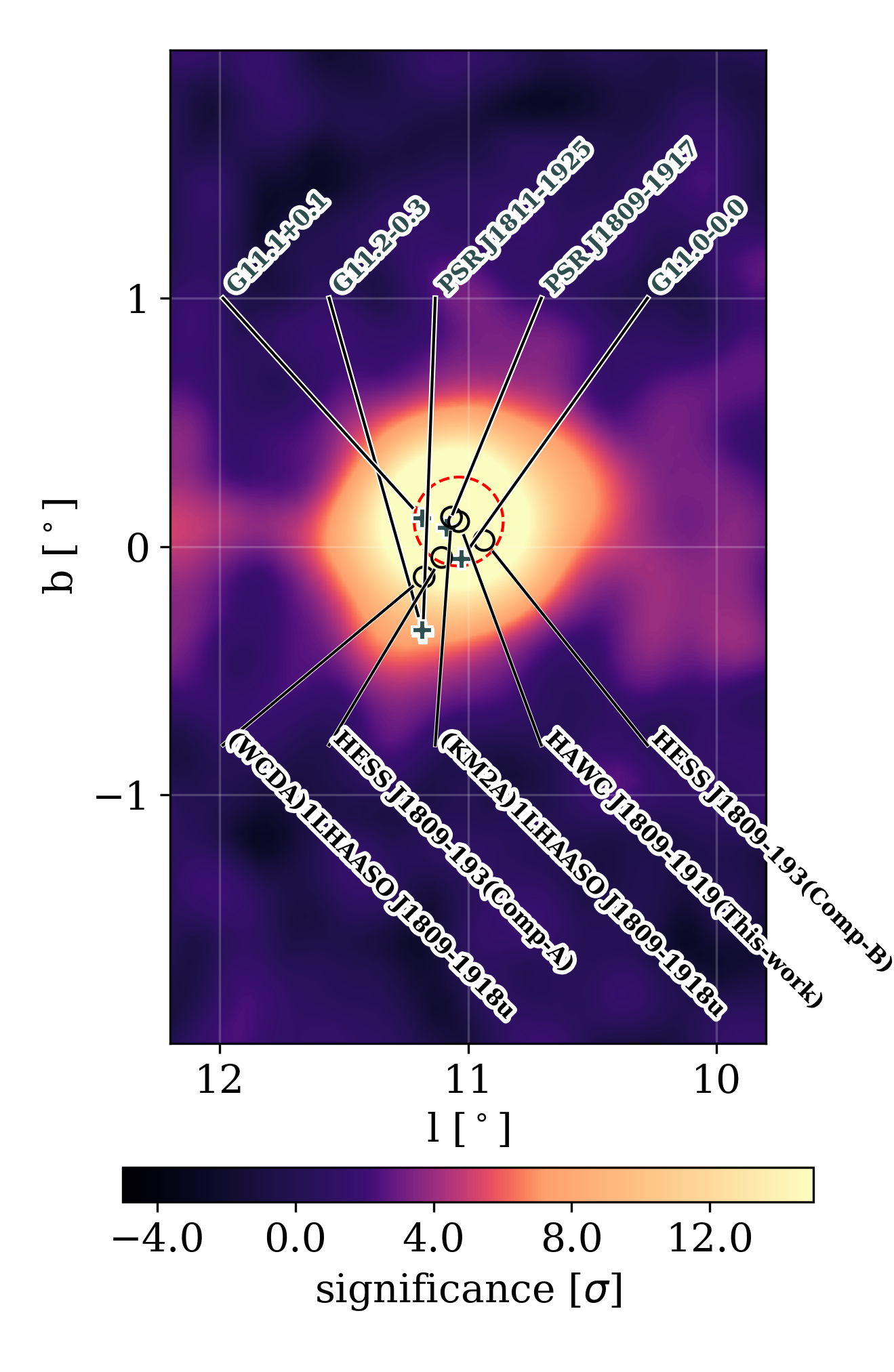}{\columnwidth}{}
\caption{HAWC sky significance map (assuming an index of 2.7 with a point source morphology) of the region shows the emission from HAWC J1809-1919. The distance between PSR J1809-1917 and HAWC fit is 0.05$^{\circ} \pm$0.28. The red dashed circle shows the 1-$\sigma$ gaussian width of the HAWC source. The maximum significance is 27$\sigma$. The upper labels show the sources in the region. The lower labels indicate the location of TeV sources observed by H.E.S.S., LHAASO, and HAWC (this work). }
\label{fig:skymap}
\end{figure}

Recently HAWC also reported that the HESS J1809-193 region is one of the few sources emitting above 56 TeV in its high-energy source catalog \citep{ehwc}. Figure \ref{fig:skymap} shows the HAWC significance map of the region using 2398 days of data, with possible counterparts within the region. It is also noted from the catalog that there is a clear indication of emission from this region above 100 TeV with significance slightly below the threshold of confirmation ($5\sigma)$. At energies above 56 TeV, the source emission remains extended with a gaussian width of 0.34$^\circ$. 
\par
The high energy emission from similar objects is particularly intriguing to study the cosmic rays near the "knee" of the cosmic ray spectrum around 1 PeV energy. The true origin of such cosmic rays is a mystery. The acceleration process of these particles continues to be a question, given that cosmic-ray accelerators produce gamma rays near their source of origin. A fraction of the energy of these cosmic rays is transferred to the gamma rays and are detected on Earth. In this work, we will explore particle acceleration in the region using a lepto-hadronic scenario and a time-dependent leptonic scenario.

\subsection{Multi-wavelength observations}

The region in the sky in the direction of HESS J1809-193 is a source-rich region, consisting of several supernova remnants (SNR) and pulsars. The radio SNR G11.0-0.05, which is a partial shell type SNR is suggested to be located at a distance of 2.4 $\pm$ 0.7 kpc\citep{shansnr}. The radio SNR G11.1+0.08 is also located within the TeV emission region. Both of these radio SNR were discovered using the Very Large Array (VLA) observations at 1465 MHz along with the Giant Metrewave Radio Telescope (GMRT) at 235 MHz \citep{snrradio}. The region also consists of a powerful high spin-down energy pulsar with a period of 82.7 ms. PSR J1809-1917 discovered by the Parkes Multibeam Pulsar Survey \citep{parkespulsar}\footnote[1]{{See https://www.atnf.csiro.au/research/pulsar/psrcat/}} is coincident with the H.E.S.S. and HAWC peak emission location. The X-ray diffuse emission discovered around SNR G11.0-0.08 \citep{xraysnr} was thought to be associated with the pulsar wind nebula(PWN) and consequently, the TeV emission was assumed to be of a PWN origin. X-ray observations from Suzaku in the 2-10 KeV band, confirmed the detection of an elongated hard, non-thermal extended emission \citep{xraysuzaku}. 
The characteristic age of PSR J1809-1917 is 51kyr \citep{parkespulsar}while the age of the SNR G11.0-0.08 is unknown which makes the association of the pulsar to SNR G11.0-0.08 difficult. Another energetic pulsar, PSR J1811-1925 is located at the eastern edge of the TeV emission, at the center of SNR G11.2-0.3 \citep{atnf}. \citet{psrj1811} mentioned that this pulsar and SNR along with X-ray binary XTE J1810-189 and the binary candidate Suzaku J1811-1900 are not responsible for the bulk of the observed TeV emission. This arises from the large location offset for PSR J1811-1925 location from the center of the emission region and XTE J1810-189 is an ordinary type I X-ray burster, and such objects have not been found to produce TeV $\gamma$-rays. 
\par
In 2016, using the Expanded Karl G. Jansky Very Large Array (JVLA), \citet{castelleti16} produced deep full-synthesis imaging at 1.4 GHz images near the vicinity of PSR J1809-1917. Along with $^{12}$CO observations from the James Clerk Maxwell telescope in the transition line J(3$\rightarrow$2) and atomic hydrogen data from the Southern Galactic Plane Survey (SGPS), a system of molecular clouds on the edge of the shock front of SNR G11.0-0.0 was discovered. This is spatially coincident with the peak emission of the source HESS J1809-193 although there are no radio counterparts detected for the PWN associated with PSR J1809-1917. They proposed that the most probable origin of the TeV emission comes from the protons of the SNR interacting with the molecular clouds in its vicinity. The density of the molecular clouds ($\sim 2-3 \times 10^3 \mathrm{cm}^{-3}$) interacting with the SNR G11.0-0.0 is found to be sufficient to produce the observed TeV gamma-ray emission in the region. As stated earlier, two unidentified LAT sources, 4FGL J1810.3-1925e and 4FGL J1811.5-1925 \citep{4fgl2} could have possible associations with the TeV source. 
\par
\section{HAWC Observatory and Description of HAWC data}
In this analysis, we use data from the HAWC Observatory to study HESS J1809-193. The HAWC detector located in the state of Puebla, Mexico at an altitude of 4100 m, consists of 300 water Cherenkov detector tanks and covers a total area of ~22000 m\textsuperscript{2}. Each tank contains four photo-multiplier tubes designed to detect Cherenkov light emitted when particles travel through water at a speed greater than the speed of light in the medium. HAWC is sensitive to sources within declinations of -26$^{\circ}$ and +64\degree and capable of continuously monitoring the sky with a total duty cycle of 95\%.
\par
This analysis uses 2398 days of data above 1 TeV energy, collected between June 2015 and June 2022 (\cite{nim}). The data is binned using a 2D binning scheme of the estimated energy and the fraction of HAWC array triggered during an event as described in \cite{crab2019}. The data is reconstructed using the neural network energy estimator\citep{crab2019}. This energy estimator algorithm uses artificial neural networks to estimate the energies of photons during an event based on the input parameters which rely on air shower characteristics such as the energy deposited by the air shower in the array, the extent of the shower footprint within the detector, and degree of attenuation of the shower by the atmosphere. The energy resolution and the angular resolution, at 10 TeV, for the neural network energy estimator at the declination of HESS J1809-193 is $\sim$15\% in logE scale and 0.4$^{\circ}$ (68\% containment radius) respectively.
\par
\section{Modeling and Results}
\subsection{Methodology}

\begin{figure}[ht!]
\fig{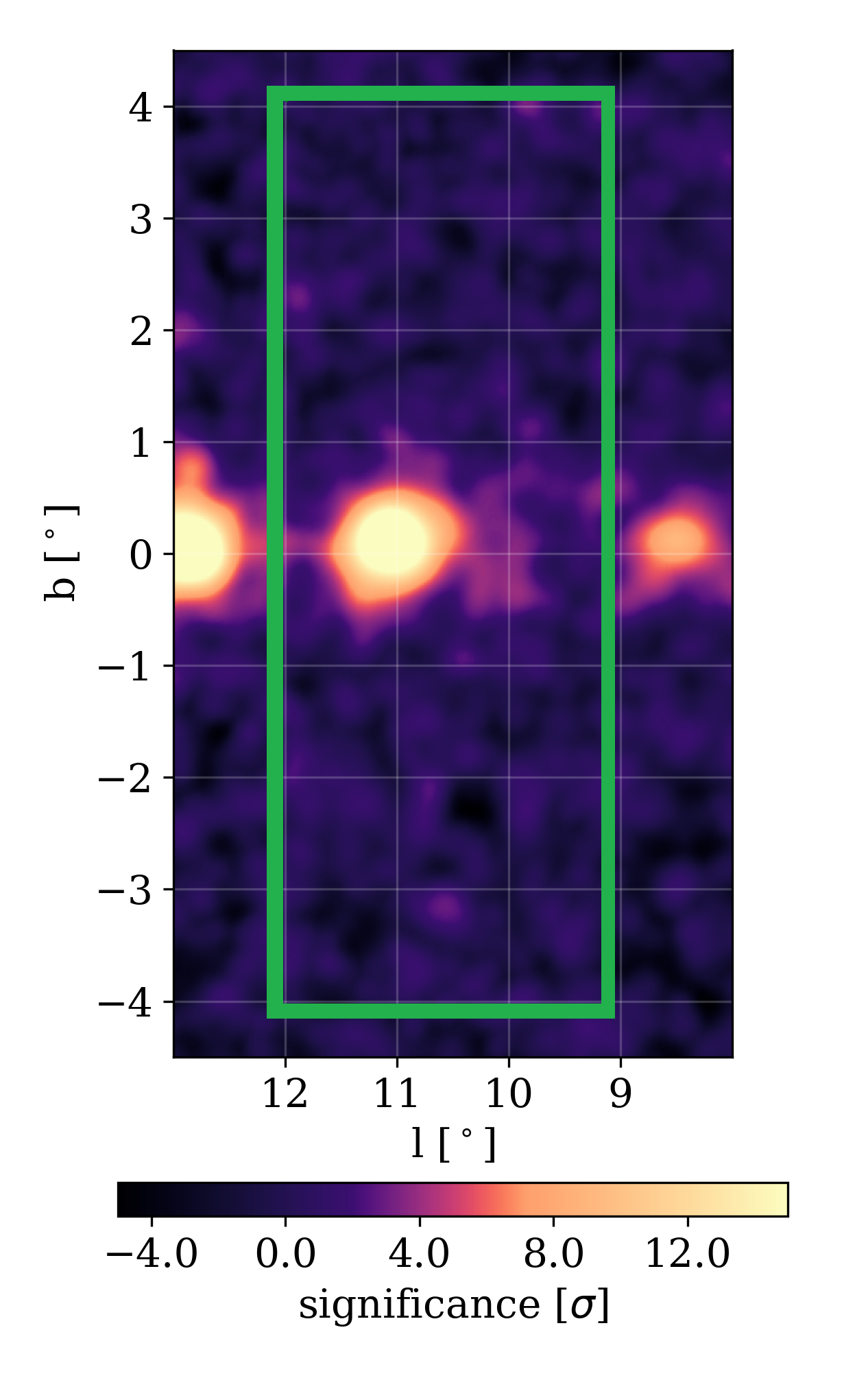}{0.9\columnwidth}{}
\caption{HAWC sky significance map(assuming an index of 2.7 with a point source morphology) of the region showing the ROI in the green rectangle.}
\label{fig:roimap}
\end{figure}

The $\gamma$-ray source morphology and spectrum are fit simultaneously with a multi-source fitting procedure using the Multi-Mission Maximum Likelihood (3ML)\footnote[2]{https://github.com/threeML/threeML} framework and HAWC Accelerated Plugin (HAL)\footnote[3]{https://github.com/threeML/hawc$\_$hal} (\citet{Vianello}, \citet{chadicrc}). The analysis is performed using a rectangular region of interest (ROI) normal to the galactic plane and defined as 9$^\circ$ $<$ l $<$ 13$^\circ$ and -4$^\circ$ $<$ b $<$ 4$^\circ$ where l, b are galactic latitude and longitude respectively. The data within the ROI, shown in Figure \ref{fig:roimap}, is used for this analysis. \par

In this analysis, we used a test statistic (TS) to calculate the pre-trial statistical significance of a model using the free model parameters. TS provides a statistical measure of how well an alternate hypothesis performs over a null hypothesis. TS is defined as:
\begin{equation}
    \mathrm{TS} = 2\ln \biggl(\frac{\mathrm{L}_{\mathrm{alt}}}{\mathrm{L}_{\mathrm{null}}}\biggl),
\end{equation}
where L$_{\mathrm{alt}}$ and L$_{\mathrm{null}}$ are likelihood values for the alternate hypothesis and null hypothesis respectively. Using Wilks’ theorem \citep{wilks}, a pre-trial significance $\sigma$, which is computed as $\sigma \simeq \sqrt{\mathrm{TS}}$, which is used to denote the significance of a source (alternate hypothesis) over background (null hypothesis). 

\par
Inspired by the Fermi-LAT Extended Source Search Catalog \citep{4FGES}, a source search method is carried out using a systematic multi-source search analysis pipeline. The pipeline to search for point sources and extended sources within the ROI is described as follows: 
\begin{enumerate}
    \item The initial phase of the pipeline analysis involves creating a TS map for the ROI assuming point-source morphology and a spectral index of 2.7. Contributions from unresolved sources and diffuse galactic background emissions, Unresolved Radiation Model (URM), are modelled using a 2D Gaussian spatial template model centered 0$^{\circ}$ along the galactic longitude. The URM model is fit to the data using 3ML and following the fitting process, the model is subtracted from the data. The resulting TS maps are checked for any remaining positive excess. 
    \item A point source model is added to the URM model at the pixel's location corresponding to the highest TS peak on the residual map. The data is then refitted using both the point source model and the diffuse template model. If the TS between this model and the previous model (URM only) is greater than 16, then this model is chosen as the new seed model, and the residual is examined for any positive excess. An additional point source model is added to the new seed model at the location of the highest TS peak in the residual, and the fitting process is repeated. This sequence is repeated until the TS of the additional point sources is less than 16 or residual TS maps no longer show significant excess (TS$<$16).  
    \item All sources, regardless of their extension, must be initially detected as a point source. Here we start checking the extension of point sources iteratively starting from the source with the maximum test statistic using the final point source model as the seed model. If the difference in the TS between the extended model and the seed model is greater than 25, then the extended model becomes the seed model for the next source extension test. Throughout the extension testing phase, if the TS of any sources becomes smaller than 16, they are removed from the model. The localization, extension and spectrum of the remaining sources are refit. This process continues until the extension tests cover all the sources inside the ROI.
    \item A similar approach, based on the source extension test, is adopted to test the spectrum of all sources. We test the spectrum of the model using a powerlaw (Eqn \ref{eq:1}), powerlaw with exponential cutoff (Eqn \ref{eq:2}) and a log-parabola spectrum (Eqn \ref{eq:3})

\begin{equation}\label{eq:1}
    \phi(E) = \phi_0 \left( \frac{E}{E_0} \right)^{-\Gamma}
\end{equation}
\begin{equation}\label{eq:2}
     \phi(E) = \phi_0\left(\frac{E}{E_0}\right)^{-\Gamma} \exp{ \left(\frac{-E}{E_{c}}\right)}
\end{equation}
\begin{equation}\label{eq:3}
    \phi(E) = \phi_0 \left( \frac{E}{E_0} \right) ^ {
          - \alpha - \beta \log{ \left( \frac{E}{E_0} \right) }
        }
\end{equation}
where $\phi_0$ is the differential flux at a pivot energy of $E_0$, $\Gamma$ is the spectral index, $E_{c}$ is the cutoff energy. $\alpha$ is the spectral index and $\beta$ is the curvature parameter for the log-parabola spectrum respectively. The pivot energy $E_0$ used in this study is chosen as 10 TeV, to minimize correlation between the spectral parameters.
\end{enumerate}

\begin{figure*}[ht!]
\centering
\gridline{\fig{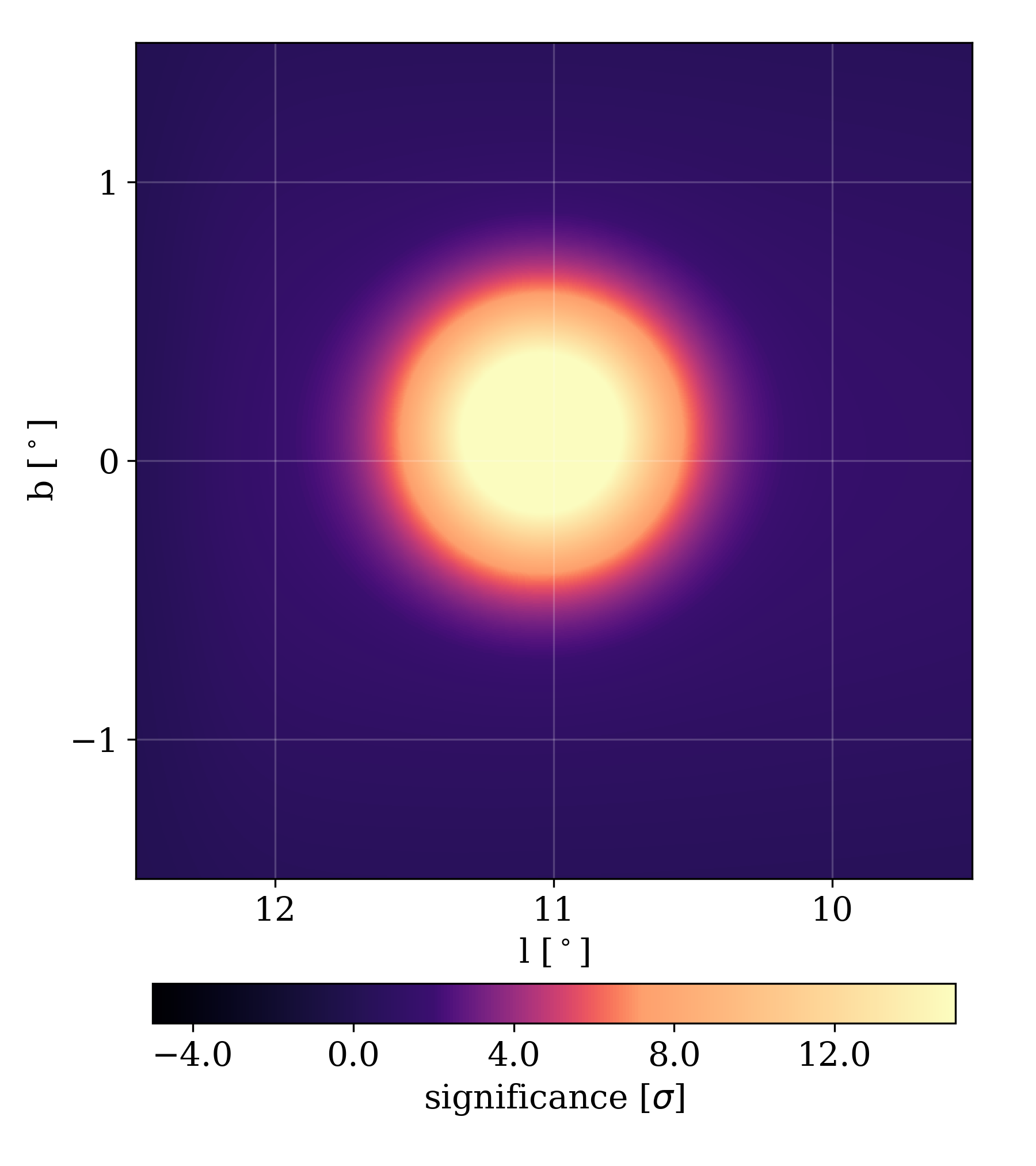}{\columnwidth}{(a)} \fig{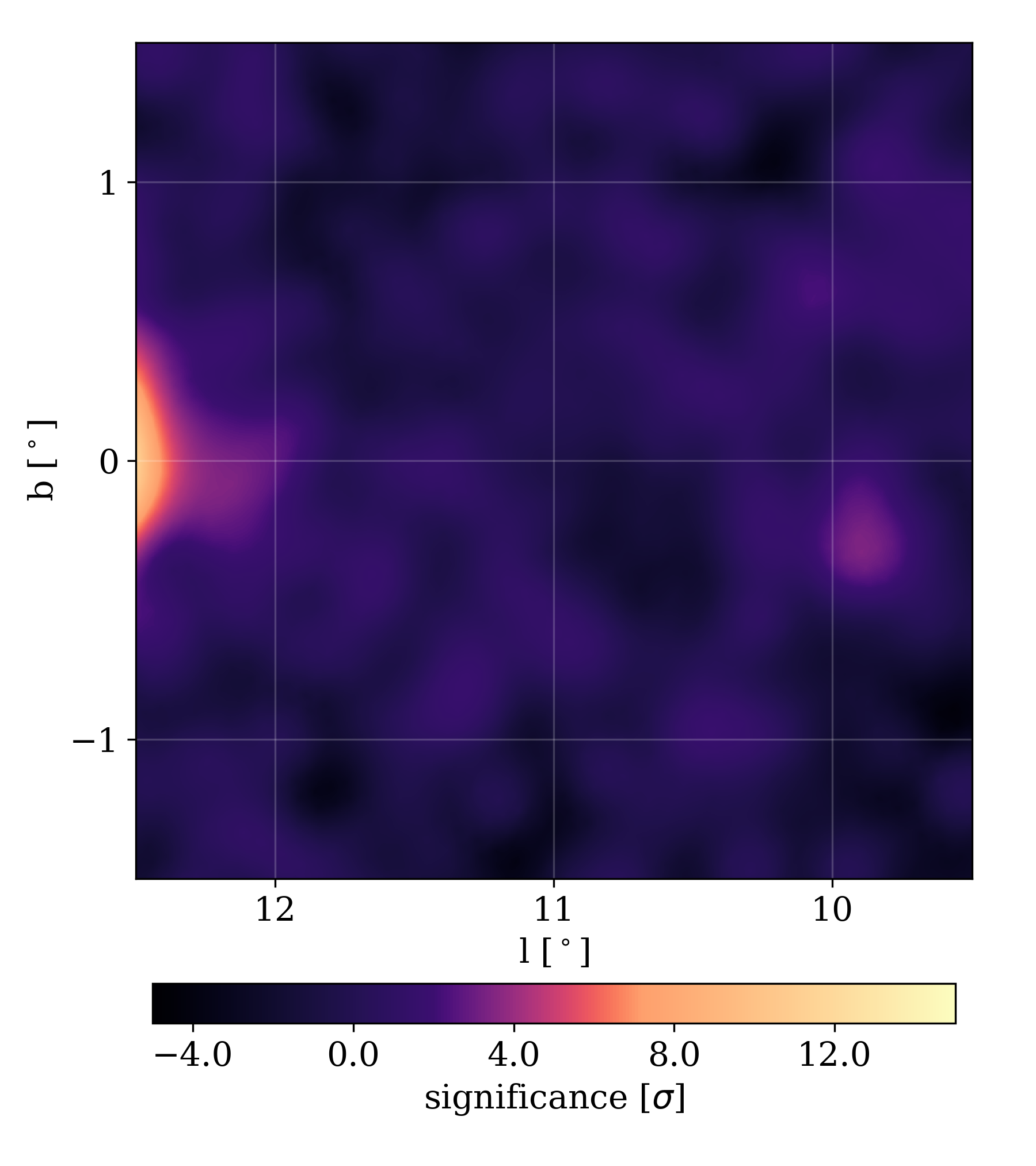}{\columnwidth}{(b)}}
\caption{(a): A significance map of the best-fit model described in Table 1. (b): A residual map produced by subtracting the best-fit model. \label{fig:modmap}}
\end{figure*}

\subsection{Results and Comparison to H.E.S.S.}
\begin{figure*}[ht!]
\fig{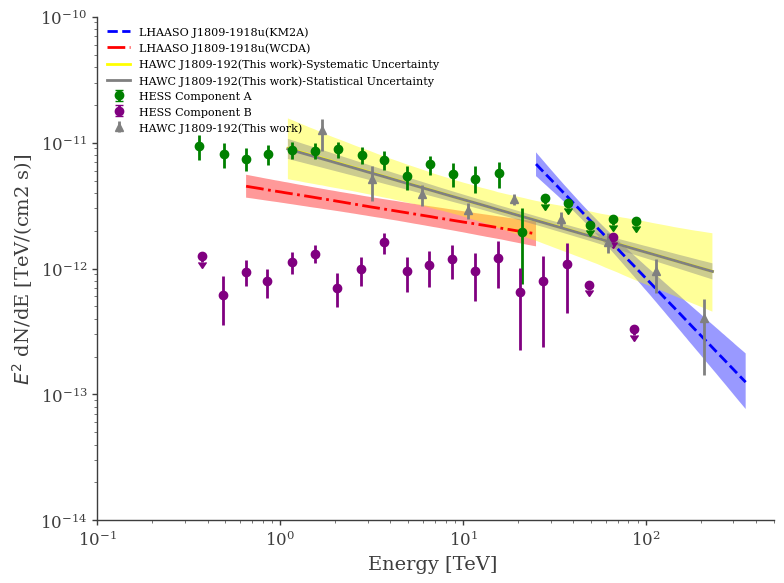}{2.0\columnwidth}{}
\caption{HAWC SED for HESS J1809-193. We compare the HAWC spectrum and flux points in gray triangles to spectra observed by H.E.S.S. \citep{new1809hesspaper} for the extended H.E.S.S. component A and the compact H.E.S.S. component B, in green and purple respectively. The LHAASO: WCDA and KM2A spectra taken from \citet{lhaasocat} are shown in red and blue respectively. HAWC spectrum (gray) shows the best-fit PL model with the statistical uncertainties. HAWC spectrum (yellow) shows the best-fit PL model with systematical uncertainties. }
\label{fig:spectrum}
\end{figure*}

The best-fit results of the source search pipeline analysis for HESS J1809-193 above 1 TeV, reveal an extended source ($\sigma$=0.21$\pm0.016_{\text{stat}}\pm0.67_{\text{sys}}$) in the region with a symmetric Gaussian morphology located at (RA, Dec) = (272.38\degree$\pm0.021_{\text{stat}}\pm0.086_{\text{sys}}$, -19.33\degree$\pm0.019_{\text{stat}}\pm0.051_{\text{sys}}$) and a power-law spectrum with an index of $\Gamma=2.42\pm0.05_{\text{stat}}\pm0.21_{\text{sys}}$. The spectral energy distribution (SED) of the source is shown in Figure \ref{fig:spectrum}) with the corresponding data points from H.E.S.S.\citep{new1809hesspaper} and LHAASO\citep{lhaasocat}. The flux points obtained from this work, show signs of a potential steepening of spectrum above 100 TeV, consistent with the spectrum observed by LHAASO above 25 TeV. Therefore we also tested the source with an ECPL model and a LGP model. We found that both complex models have a similar statistical significance, $\sim 1 \sigma$ and $\sim 1.2 \sigma$ respectively compared to the power-law, and are not preferred over the power-law model by $3 \sigma$ significance. Therefore we adopted the simpler power-law model to best describe the region. More data at higher energies is needed to make conclusive evidence for the curvature of the spectrum. 
\par

Figure \ref{fig:modmap}(a) shows the model map for the region and Figure \ref{fig:modmap}(b) shows the residual map after the subtraction of model from the data map. Table \ref{bestfitvalue} shows the best-fit parameters from modeling the source at a pivot energy of 10TeV, along with their statistical and systematic uncertainties \citep{crab2019}. Systematic uncertainties related to the modeling of the HAWC detector are investigated as described in \citet{crab2019}. The effects of the detector systematic uncertainties are shown as the yellow band in Figure \ref{fig:spectrum}. The energy range for this source is determined by adding a step function cutoff to the best-fit model, and calculating the maximum and minimum energies with a 1$\sigma$ confidence level. 
\par
The analysis was repeated using data above 56 TeV. We find that the source becomes softer above 56 TeV, with a best-fit index of $\Gamma$ = 2.94 $\pm$ 0.29$_{\mathrm{stat}}$, and remains extended with a symmetric Gaussian morphology with a 1-$\sigma$ extension of 0.186$^{\circ} \pm$ 0.023$_{\mathrm{stat}}$. An energy dependent morphology study for HESS J1809-193 based on the methodology explained in Section 4. of \citet{j2019} was done. The results of the energy-dependent morphology lacks conclusive evidence for change in morphology with increasing energy, primarily due to the large uncertainties in the measurements attributing from poor angular resolution of HAWC at the declination of the source. This may be improved through additional data in the future, along with HAWCs outrigger array\citep{outriggericrc}.

\par

In contrast to the detection of a two-source model from H.E.S.S. \citep{new1809hesspaper} (with an angular resolution of \textless0.1$^{\circ}$ above 1 TeV), HAWC (with an angular resolution of 0.55$^{\circ}$ (68\% containment radius) above 1 TeV at the declination of HESS J1809-193), detects a single extended source which is likely due to the different energy ranges and angular resolution of the two instruments. The comparison of spectral results suggests that the spectrum of the observed HAWC source is similar to that of the more brighter component, Component A, as detected by H.E.S.S.

\begin{table}[h!]
\renewcommand{\arraystretch}{1.4}
  \centering
  \begin{tabular}{cc}
  \hline
    Fit parameters & Best fit values\\
    \hline
    TS & 392 \\
    RA(\degree) & 272.38$\pm0.021_{\text{stat}}\pm0.086_{\text{sys}}$ \\
    Dec(\degree) & -19.33$\pm0.019_{\text{stat}}\pm0.051_{\text{sys}}$ \\
    $\phi_0$(TeV$^{-1}$ cm$^{-2}$ s$^{-1}$) & 3.53$\pm0.31_{\text{stat}}\pm1.19_{\text{sys}} \times 10^{-14}$ \\
    Index & 2.42$\pm0.05_{\text{stat}}\pm0.21_{\text{sys}}$ \\
    $\sigma$(\degree) & 0.21$\pm0.016_{\text{stat}}\pm0.67_{\text{sys}}$ \\
    \hline
  \end{tabular}
  \caption{Best-fit parameters and their statistical and systematic uncertainties for HAWC J1809-1919. $\phi_{0}$ is the flux normalization at a pivot energy of 10 TeV. Also given is the 1$\sigma$ radius of the Gaussian morphological model.}
\label{bestfitvalue}
\end{table}
\section{Spectral Modeling}
Due to the number of SNRs along with potential PWNs in the region, it is unclear whether the observed 
$\gamma$-rays have a hadronic or leptonic origin. Studies performed by \citet{castelleti16} using \textsuperscript{12}CO data in the region of HESS J1809-193 revealed a system of molecular clouds positionally coincident with the peak of the observed $\gamma$-ray emission from the H.E.S.S. observations. X-ray observations by Suzaku \citep{xraysuzaku} reveal a hard non-thermal spectrum(photon index $\Gamma \approx 1.7$). Therefore in section 4.1 we investigate the lepto-hadronic scenario for the multi-wavelength observations from the contribution of SNR-molecular gas cloud interaction and the pulsar PSR J1809-1917 injecting particles into the system using NAIMA \citep{naima}framework. In section 4.2, we investigate a time-dependent leptonic scenario for the PWN associated with PSR J1809-1917 using GAMERA \citep{gamera} framework. We also note that the associations made to the multiwavelength data are used to explain the SED alone and not the spatial features.

\subsection{Lepto-hadronic scenario}
\begin{figure*}[ht!]
\fig{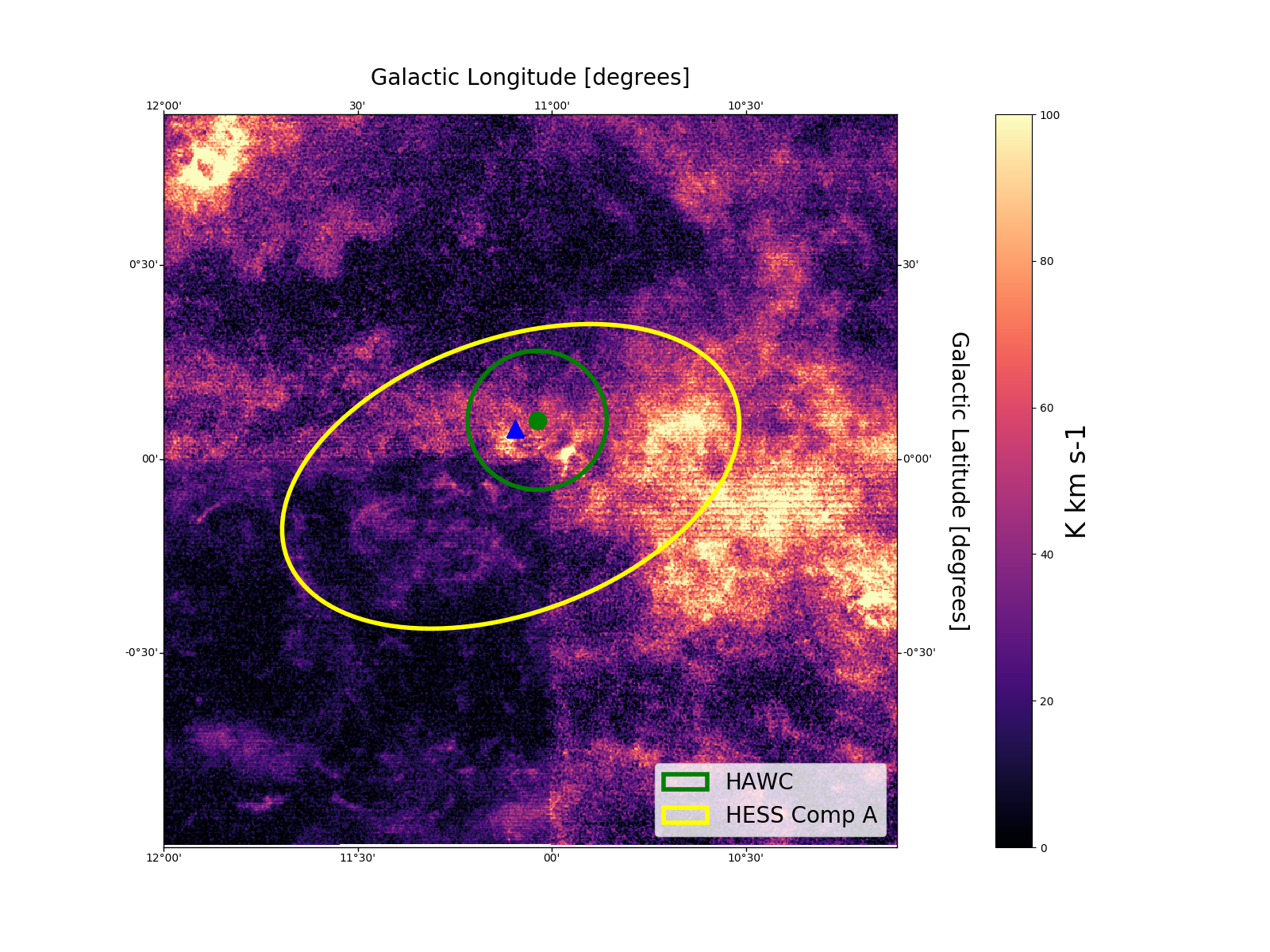}{1.8\columnwidth}{}
\caption{$^{12}$CO (J=1-0) FUGIN line emission map \citep{fugin} for the HESS J1809-193 region. The blue triangle represents the location of PSR J1809-1917, the green dot and circle represent the HAWC source position and the 1-$\sigma$ extension, and the yellow square and ellipse represent the H.E.S.S. extended source respectively. }
\label{fig:fugin}
\end{figure*}

Observations of molecular clouds by \citet{castelleti16} and \citet{viosin} along with the presence of SNRs, increase the probability of cosmic rays accelerating in the SNR shocks and interacting with the molecular clouds in the region. Figure \ref{fig:fugin} shows the $^{12}$CO (J=1-0) FUGIN line emission contours overlaid with the HAWC and H.E.S.S. extensions that show the morphology of the molecular clouds detected within the region. The distance estimates for SNR G011.0-0.0, as reported in the studies, include approximately 3.7 kpc according to \citet{viosin}, around 3.0 kpc based on \citet{castelleti16}, 2.4$\pm$0.7 according to \citet{shansnr}, and 2.6 kpc according to \citet{xraysnr}. These estimates are closer to the distance estimations for PSR J1809-1917, which is approximately 3.7 kpc. The distance estimate for SNR G011.1+00.1 is 17 kpc, with a 40\% fractional error \citep{snrg111}. This would, unrealistically, place SNR G011.1+00.1 outside the boundaries of the galaxy. Hence for this study, we do not consider the emissions from SNR G011.1+00.1. Since SNR G011.0-0.0 is believed to be the progenitor of PSR J1809-1917, we show the integrated radio flux points from SNR G011.0-0.0 calculated by \citet{brogan}, for comparison in the multiwavelength analysis. 

H.E.S.S. detection of two components along with the break in the spectra in the GeV energy range motivates using a complex model, involving contributions from both leptonic and hadronic models(lepto-hadronic model), over a simple hadronic model. This model assumes the SNR/molecular cloud interaction for the hadronic mechanism motivated by the overlap of the HAWC and the molecular cloud morphologies from Figure \ref{fig:fugin} (no spatial information is used in the study). For the leptonic scenario, we assume the electrons injected into the PWN by the pulsar, PSR J1809-1917. For this physical scenario, we use NAIMA to model the multi-wavelength observations. 
\par
The very high-energy $\gamma$-rays observations from HAWC and the H.E.S.S. extended source component(Component A) are modelled with a hadronic population. Even though the molecular cloud density in the region shows a gradient \citep{castelleti16}, seen in Figure \ref{fig:fugin}, the TeV morphology does not show any similar characteristics across the region of HAWC J1809-1919. Therefore we calculate an average ambient density that can produce the detected  TeV photons from the region. To quantify the ambient density within the region, we first calculate the $\gamma$-ray flux for HAWC J1809-1919 above 1 TeV by taking into account the power-law spectrum of this source (d$\phi_\gamma$(E)/dE=$\phi_0$(E/10 TeV)$^{-\Gamma}$) with $\phi_0$ and $\Gamma$ values from Table(\ref{bestfitvalue}), which implies a F$_\gamma(> 1\text{TeV}) = 6.4 \times 10^{-12} \text{ph cm}^{-2} \text{s}^{-1}$. Assuming the fraction of the total supernova explosion
energy converted to cosmic-ray energies, $\theta$=10\%  \citep{snreff}, the distance to the SNR, d$\sim$3.5 kpc and the supernova energy output of $10^{51}$ erg, we calculate the required ambient density in the region to produce the observed $\gamma$-ray emission, using Eq.(16) of \citet{gammaeqn}.  We find that a minimum ambient density of $n\sim40 \text{cm}^{-3}$ at a distance of 3.5 kpc, is required to produce the $\gamma$-ray emission detected above 1 TeV. This calculated value is below the density of cloud in the range of 2-3$\times10^3 \mathrm{cm}^{-3}$estimated by \citet{castelleti16}, using 12CO observations.

\par
The Suzaku X-ray observations are modelled using a leptonic population and the radio observations are used only for comparison. The X-ray flux is measured in the vicinity of the nebula(regions 2, 3, 6 and 7 in Table 4 of \citet{xraysuzaku}). The same leptonic population can produce TeV emissions through the Inverse Compton (IC) mechanism(assuming cosmic microwave background (CMB), Far-infrared radiation (FIR), and Near-infrared radiation (NIR) photon fields as seed photons with their values set to (2.72 K, 0.261 eV/cm$^3$), (30 K, 0.5 eV/cm$^3$) and (3000 K, 1 eV/cm$^3$) respectively, obtained from GALPROP). We also consider the same leptonic population interacting with the molecular cloud (with a ambient ion density n$_{\text{ion}} \sim 40$  cm$^{-3} $) undergoing a non-thermal Bremmstrahlung process to reproduce the observed Fermi data points obtained by \citet{new1809hesspaper}. The parent particle spectrum is assumed to follow an exponential cut-off power-law, motivated by the spectrum break between LHASSO KM2A and WCDA (used only for comparison), and with an assumed distance of 3.5 kpc distance to the source. 
\par
\begin{figure*}[ht!]
\gridline{\fig{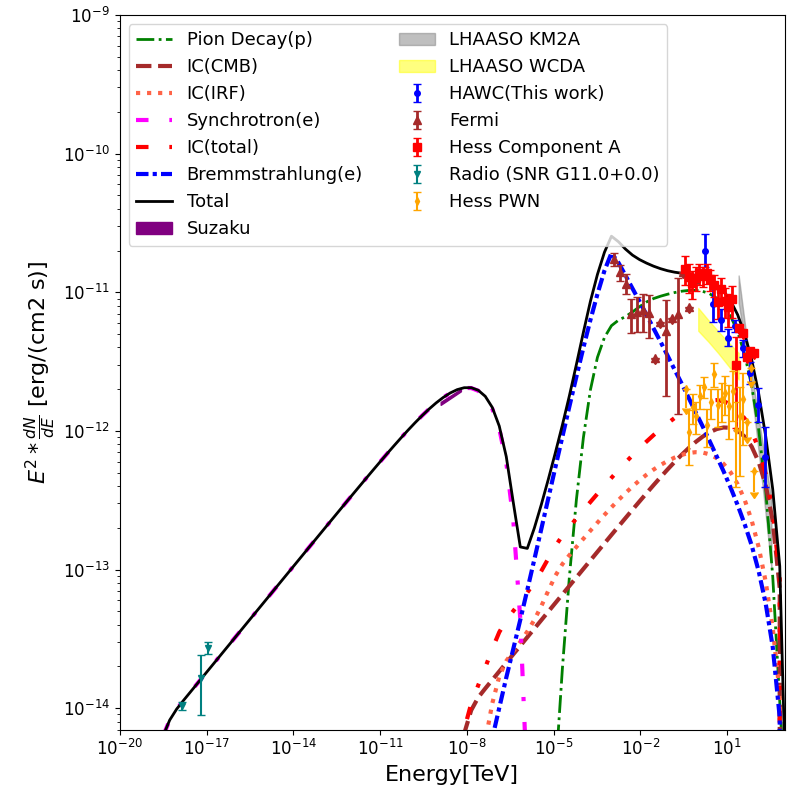}{\columnwidth}{(a)Full SED}\fig{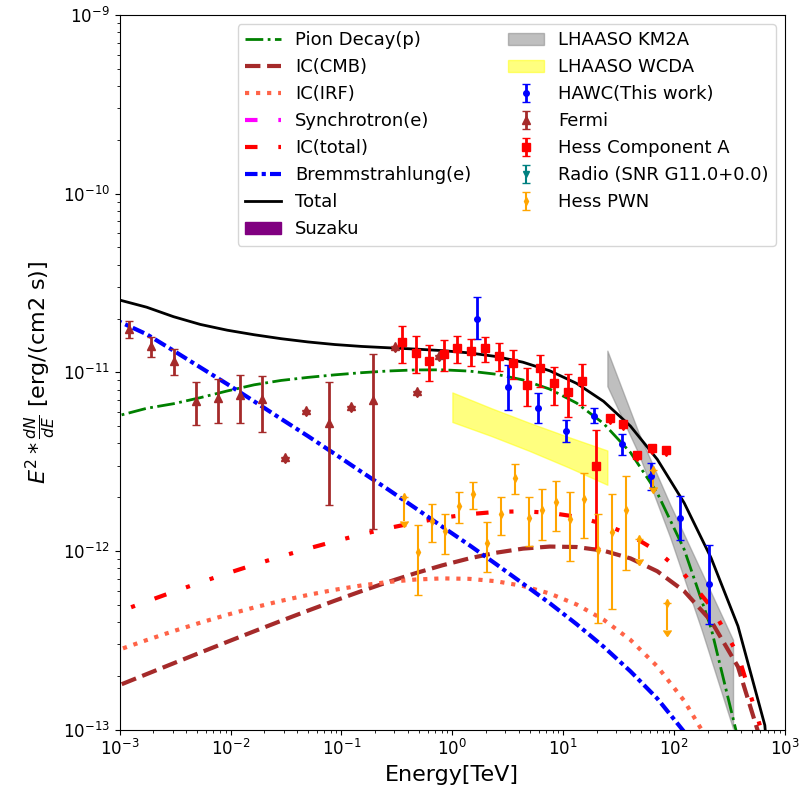}{\columnwidth}{(b)Zoomed-in $\gamma$-ray SED}}
\caption{Multi-wavelength SEDs for the lepto-hadronic scenario of HESS J1809-193 region obtained with NAIMA. The data shown includes the 90 cm radio observations of the SNR G11.0-0.0 with the Very Large Array survey points in teal \citep{snrradio}, the Suzaku butterfly spectra in purple \citep{xraysuzaku}, the Fermi-LAT, H.E.S.S. Component A and H.E.S.S. Component B flux points, in brown, orange and green respectively \citep{new1809hesspaper}, and the HAWC flux points from this work in blue. Figure(a) shows the full SED for the region. The primary synchrotron is shown in dashed pink line. Total inverse compton SED is show in dashed red line while the individual seed photon components for IC: CMB and interstellar radiation fields (IRF) are shown in dashed brown and orange tomato colors respectively. p-p emission is shown in green and the non-thermal Bremsstrahlung emission is shown in blue dashed lines.The total SED is represented in gray. Figure(b): Zoomed in version of the SED for $\gamma$-ray observations. Shown for comparison, but not used in the fit, is the spectrum for the ultra high-energy source 1LHAASO J1809-1918u from the \citep{lhaasocat} and the radio flux points from SNR G11.0+0.0}
\label{fig:spectralmodel}
\end{figure*}

The model parameters are fitted to the multiwavelength data using the Markov Chain Monte-Carlo (MCMC) procedure to obtain probability distributions for the parameters of the parent particle populations. The fit parameters for the model are summarised in Table \ref{tab:naimafit} and the resulting spectra are shown in Figure \ref{fig:spectralmodel}. 
Figure \ref{fig:spectralmodel}(a) shows the total SED for the resultant model, while Figure \ref{fig:spectralmodel}(b) shows the SED in the GeV-TeV. LHAASO WCDA and KM2A spectra from LHAASO's catalog \citep{lhaasocat} are also plotted for comparison. We observe the total spectrum can reproduce all the observed emissions including the highest energy HAWC and LHAASO data.
\par

 We tested the hadronic component in Naima with two different maximum proton energies, E$_p$, at 800 TeV and 1 PeV, and found that a proton energy of at least 1 PeV is required to explain the TeV $\gamma$-rays. The hadronic component prefers a relatively hard index of $\Gamma_{p} = 2.00\pm 0.03$ with a cutoff of 350 TeV. Integrating the spectrum above 1 GeV gives a total proton energy of $W_{p} \sim 1.23 \times 10^{49} \text{erg}$, which is a very small fraction of the calculated energy released in a supernova explosion of $\sim 10^{51}$ erg \citep{snrenergy}. Considering the gas density gradient across the HAWC 1$\sigma$ width of 0.21$^{\circ}$ observed by FUGIN(Figure \ref{fig:fugin}), cosmic rays accelerated by SNR G011.0-0.0 can potentially reach the dense molecular clouds in the region ($n_h \sim 40\ \mathrm{cm}^{-3}$) and produce the observed $\gamma$-rays. In the case of the proton-proton interaction scenario, the presence of secondary synchrotron component from the secondary electrons produced in the process is expected. Observations of the region with larger field-of-view with X-ray telescopes would provide evidence of such an emission.
 
 The secondary electrons in the pp scenario is not adding complexity to the model, they are just expected to be present. They need to be mentioned even their synchrotron emission is not computed.

 \par
 
 For the leptonic model, we get an index of $\Gamma_{e} = 2.49\pm 0.06$ and a relatively low magnetic field of B=2.8 $\pm 0.11\mathrm{\mu}$G. The synchrotron emission from the leptonic model shows good agreement with the Suzaku X-ray data. We also observe that this same population of electrons at higher energies can reproduce the spectrum of the compact component detected by H.E.S.S. (H.E.S.S. Component B) through the IC mechanism. This could explain the X-ray component along with H.E.S.S. component B, and is compatible with the emission surrounding the PWN associated with PSR J1809-1917. The same leptonic population could interact with the molecular cloud in the vicinity undergoing a Bremmstrahlung mechanism to reproduce the $\leq$100 GeV $\gamma$-ray emission observed by Fermi.

\subsection{Time-dependent leptonic model}

In this section, we explain the spectra assuming a leptonic scenario, where the emission originates from a PWN powered by the pulsar PSR J1809-1917. The observed period ($P$) and the period time derivative ($\dot{P}$) are 82.7 ms and $2.553 \times 10^{-14}$\citep{atnf}. The characteristic age, $\tau_{c}$, of the pulsar is 51.3 kyr which is an estimated measure for the pulsar age, provided the assumption of pulsar braking index n=3 and the birth period ($P_{0}$) is less than the observed period $P$ holds true \citep{gamerapwnevol}. The distance used to calculate the flux in this model is assumed to be 3.5 kpc. We found that a model with a minimum of three particle populations is required describe the observed TeV and X-ray data. The radio flux data is shown only for comparison and is not included in the fit. In this model, the fraction of the pulsar spin-down luminosity converted to electrons ($\theta$-fraction), particle cut-off energy ($E_c$), and the birth period ($P_0$) are treated as unknown parameters used to describe the injected electron spectrum. The true age of the system,$\tau$, is calculated as a function of $P_0$ using equation(\ref{eq4}). 

\begin{equation}\label{eq4}
    \tau = \frac{P}{(n-1)\dot{P}}\left[ 1-\left(\frac{P_0}{P}\right)^{n-1}\right]
\end{equation}
\par
\par
We use the GAMERA package\citep{gamera} to model the spectrum of the radiation produced by the particles. GAMERA can produce a time-dependent model of relativistic electrons, including its injection and cooling, producing photon emission in different wavelengths. In this model we consider a power-law with an exponential cutoff for the injected electrons. The seed radiation fields used to calculate the IC spectrum are calculated from \citet{radfield} for the location of PSR J1809-1917. \citep{gamerapwnevol} specifies the conditions for the time evolution for $B$, $\dot{E}$ and $P$. The time-dependent modelling approach is similar to the approach outlined in \citet{j2019} and \citet{new1809hesspaper}. The parameters of the model are described in Table(\ref{tab:gamerafit}). 

\par

We fit the model to three generations of electrons: relic, middle-age and recently injected `young electrons'. Relic electrons are electrons injected into the ISM by the PWN over the lifetime of the system, in this case the result of the fit gives the age of the system as $\tau \sim $26.5 kyr, and is associated with the HAWC, H.E.S.S. Component A and LHAASO WCDA components. Medium-age electrons are electrons injected into the system in the last 10.2 kyr and associated with the compact H.E.S.S. Component B. Young electrons are electrons injected into the system in the last $\sim$3.1 kyr and is associated with the X-ray nebula. We also see that the evolved spectra has an index of $\Gamma$=2.1 and an evolved present day magnetic field, B$\sim$3.2 $\mu$G and about 67\% of the pulsars spin-down luminosity is converted to electrons. This signifies that the spin-down energy of the pulsar is sufficient to maintain the energy of electrons and positrons in the wind of the pulsar powering the entire multiwavelength observations.


\begin{figure*}[ht!]
\gridline{\fig{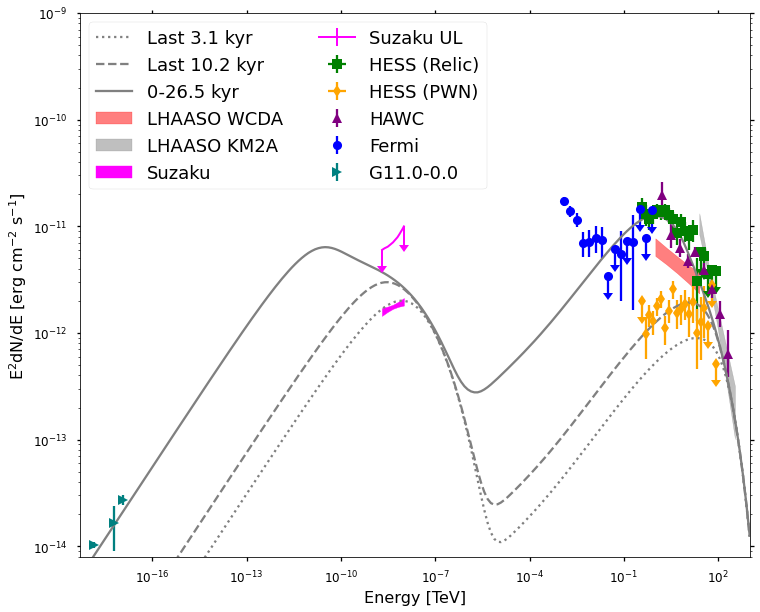}{1.05\columnwidth}{(a)Full SED}\fig{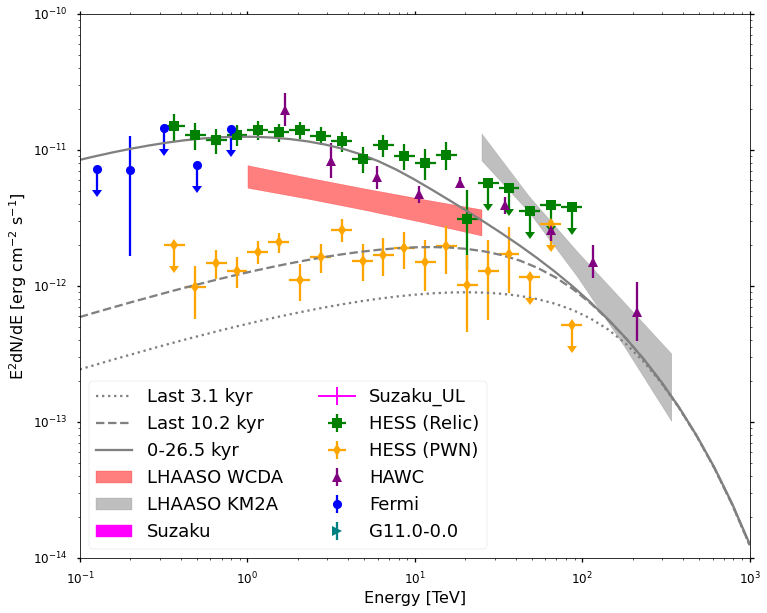}{1.05\columnwidth}{(b)$\gamma$-ray SED}}
\caption{Multiwavelength SED for the time dependent leptonic model of HESS J1809-193 obtained with GAMERA. The presented data is identical to the one shown in Fig(\ref{fig:spectralmodel}). Fig(a) shows the full SED for the region for the three generations of electrons. The solid line indicates the old 'relic' electrons. The dashed lines and curved lines correspond to the medium-age and young electrons injected into the system respectively. Fig(b) shows the zoomed-in $\gamma$-ray observations and their MCMC results. Shown for comparison(but not used in the fit is the spectrum for the ultra high-energy source 1LHAASO J1809-1918u from  \citet{lhaasocat}.}
\label{fig:gamera}
\end{figure*}

\begin{table}[h!]
\renewcommand{\arraystretch}{1.2}
  \centering
  \begin{tabular}{cc}
  \hline
    Parameters & Fit values\\
    \hline
     \multicolumn{2}{c}{Leptonic fit} \\
     \hline
    log($A$[1/TeV]) & 43.81 $\pm$ 1.14  \\
    $E_0$[TeV] & 10  \\
    $\Gamma_e$ & 2.49 $\pm$ 0.06 \\
    $E_{\text{cutoff}}$[TeV] & 630.54 $\pm$ 185 \\
    B-field[$\mu$G] & 2.80 $\pm$ 0.11  \\
    W $>$ 1 GeV[erg] & $2.35^{+1.19}_{-0.95}{} \times 10^{48}$ \\
    \hline
        \multicolumn{2}{c}{Hadronic} \\
    \hline
    log($A$[1/TeV])  & 45.91 $\pm$ 1.17 \\
    $E_0$[TeV] & 10 \\
    $\Gamma_p$  & 2.00 $\pm$ 0.03\\
    $E_{\text{cutoff}}$[TeV] & 349.05 $\pm$ 65.33\\
    W $>$ 1 GeV[erg] & $1.23^{+0.09}_{-0.1}{} \times 10^{49}$ \\
    \hline
  \end{tabular}
  \caption{Naima best-fit parameters for the spectral models shown in Figure \ref{fig:spectralmodel}. Note that $E_0$ is fixed here.}
  \label{tab:naimafit}
\end{table}

\begin{table}[h!]
\renewcommand{\arraystretch}{1.2}
  \centering
  \begin{tabular}{ccc}
  \hline
    Parameter & Parameter description & Fit values\\
    \hline
     & \text{Pulsar Parameters} & \\
     \hline
$\dot{\text{E}}$ & \text{Spin down power}  & $1.8 \times 10^{36}\ \text{erg s}^{-1}$ \\ 
 $\tau_c$ & \text{Characteristic age} & 51\ \text{kyr} \\
 P & \text{Pulsar period} & 82.76\ \text{ms} \\
 $\dot{\text{P}}$ & \text{Pulsar period derivative} & $2.55 \times 10^{-14} \text{s s}^{-1}$ \\
 d & \text{Distance to pulsar} & 3.3\ \text{kpc} \\
 n & \text{Braking index} & 3 \\
 \hline
  & \text{Fitting parameters} & \\
  \hline
 $\theta$ & \text{Power fraction}  & 0.67 \\ 
 $\text{E}_{c}$ & \text{Particle Cut-off energy} & 794\ \text{TeV} \\
 $\alpha$ & \text{Particle injection index} & 2.1 \\
 B & \text{Magnetic field} & $3.2\ \mu\text{G}$ \\
 $\text{P}_0$ & \text{Pulsar birth period} & $47\ \text{ms}$ \\
 $\tau_{\text{young}}$ & \text{Age of young electrons} & 3.1\ \text{kyr} \\
 $\tau_{\text{med}}$ & \text{Age of middle-age electrons} & 10.2\ \text{kyr} \\
 $\tau_{\text{relic}}$ & \text{Age of relic electrons} & 26.5\ \text{kyr} \\
 \hline
  \end{tabular}
  \caption{Pulsar parameters and fit parameters for the time-dependent PWN model. The listed values are the Pulsar parameters taken from ATNF catalog \citep{atnf} and the fit parameters obtained from an MCMC fit to the spectral data using GAMERA.}
\label{tab:gamerafit}
\end{table}

\section{Discussion}
The outcome from the SED modelling and analysis suggest two potential models to describe the observed multiwavelength spectra: (i)a lepto-hadronic scenario, describing the potential involvement of CRs produced by the SNR and PWN, (ii) a time-dependent model, with 3 population of electrons injected into the ISM by the PWN system, each associated with their respective emission wavelengths (TeV $\gamma$-ray and X-ray). 

In the lepto-hadronic scenario for the observed multiwavelength SED, the protons accelerated at the shocks from SNR G11.0-0.0, SNR G11.1+0.1 or both, could be responsible for the observed GeV-TeV $\gamma$-ray emission. While the age of these SNRs remains unclear, measurements done based on the association of 12CO and HI-self absorption features done by \citet{castelleti16} places the SNR G11.0-0.0 at approximately 3 kpc, within the uncertainties of the derived kinematic distance. This could indicate the association of SNR G11.0-0.0 as the progenitor of PSR J1809-1917. Protons accelerated by SNR G11.0-0.0 could interact with the molecular gas clouds in their immediate vicinity to produce the observed TeV emission. The total mass and total proton density , as measured by \citet{castelleti16}(taking into account contributions from both HI and $^{12}$CO emissions), from the molecular clouds estimated to be $\text{M}\approx3 \times 10^3 M_{\odot}$ and $\eta \approx 7.4 \times 10^{3}\ \mathrm{cm}^{-3} $, is found to fulfill the required amount of target material for hadronic interactions. We conclude that a proton energy of at least 1 PeV is required for a distance of 3.5 kpc to the source and an average molecular cloud density for proton interactions of $\sim$40 cm$^{-3}$ is required within the region of high-energy $\gamma$-ray emissions for a SNR shock scenario. This could indicate that the source is a hadronic PeVatron. The estimated value of the present-day B-field is $\sim 2.8\ \mu$G, which is low relative to the ISM and it depends on the normalization of the peak of X-ray synchrotron emission. While the magnetic field is comparatively low, the synchrotron emission can explain the X-ray data. Considering the X-ray emission is produced near the PWN, the same population of electrons is also able to reproduce the TeV emission observed from the compact H.E.S.S. Component B, which spatially coincident with the pulsar and the PWN. 
\par
For the time dependent PWN scenario, we show SEDs obtained for three generations of electrons. The model describes well the spectra of HAWC and H.E.S.S. Component A and the X-ray flux from Suzaku. The fit yields an average present day magnetic field of $\sim 3.2 \mu$G and a reasonable index for the injection spectrum of $\sim 2.1$. A maximum electron energy of several hundred TeV is also observed from the spectrum fit, which is similar to the maximum electron energy observed in the lepto-hadronic fit. The model predicts that the 'relic' old electrons injected into the system would cool down over time, which could explain the slight cutoff observed at the highest energies. In the case of the relic electrons, we find that the observed total synchrotron emission overshoots the observed SED by a factor of 2-3 at energies between 2-10 keV. This is potentially due to the smaller region of interest ($\sim 0.3^\circ \times 0.3^\circ$) for the Suzaku observations. \cite{xraysuzaku} found diffuse X-ray emission extending beyond south of the PWN. Therefore if the origin of the TeV $\gamma$-rays is of leptonic origin, then a diffuse X-ray emission from synchrotron radiation could surround the X-ray PWN with a morphology similar to the observed TeV one. Therefore to quantify the flux from this diffuse X-ray region, \citet{new1809hesspaper} calculated an upper limit (95\% CL) for the X-ray flux shown in Figure(\ref{fig:gamera}). \citet{new1809hesspaper} observed in their modeling that the synchrotron emission from relic electrons match the upper limit. Our modeling results show that the synchrotron flux produced by both relic and middle-age electrons are well below the X-ray upper-limits at keV energies.
\par
This model scenario however cannot explain the Fermi-LAT observations below $\sim$10 GeV. This could be attributed to a secondary population of particles, undergoing a non-thermal Bremsstrahlung process near the pulsar with the gas density in the region as mentioned previously. 

\section{Conclusions and Outlook}
In this work, we perform a detailed spectral and morphological study of HESS J1809-193 using 2398 days of HAWC data. Unlike the H.E.S.S. observations for this region, HAWC observations indicate a single extended source in the region and no conclusive evidence for energy dependent morphology due to the low angular resolution of the detector at higher zenith angles. The morphology of the region reveals an extended symmetric gaussian source with a 1-$\sigma$ extension of 0.21$\pm0.016_{\text{stat}}$ (above 1 TeV) and 0.186$^{\circ} \pm$ 0.023$_{\mathrm{stat}}$ (above 56 TeV). Spectral studies indicate a power-law spectrum with an index of $\sim$2.44. There could be a potential steepening of the spectrum at the highest energies indicated by the last flux point above 200 TeV, which requires more data/statistics. The HAWC spectrum extends the emission observed by H.E.S.S. well past 56 TeV.

\par
Studies done by \citet{new1809hesspaper}, have considered a time-dependent leptonic scenario to explain the multiwavelength spectrum of HESS J1809-193 by invoking the electrons injected into the system by PSR J1809-1917. A lepto-hadronic model proposed by \citet{boxi}, was also used to explain the emission of HESS J1809-193, where the explosion of SNR G11.0-0.0 injected CR protons and electrons into the region producing the high-energy gamma-rays. In our studies, we have expanded on the multiwavelength model proposed by HESS using newer HAWC data and place better constraints on the model. We also suggested that a lepton-hadronic scenario involving the SNR injecting CR protons and the PWN produced by PSR J1809-1917 injecting CR electrons into the system, can better explain the observed high-energy gamma-rays and X-rays simultaneously, except for the Fermi-LAT data below $\sim$ 10 GeV.

As mentioned earlier, this work is based on spectral association alone and the detailed study of spatial associations is left for future studies. Hence we cannot provide conclusive evidence for the origin of the $\gamma$-rays from the region of HESS J1809-193. Conclusive evidence of $\gamma$-ray production from the region requires deep morphological and spectral studies of the SNRs present. Future observations from LHAASO, the upcoming Cherenkov Telescope Array (CTA), and the Southern Wide-Field Gamma-Ray Observatory (SWGO) could provide unprecedented observations to enhance our understanding of the region.
\section{Acknowledgements}
We acknowledge the support from: the US National Science Foundation (NSF); the US Department of Energy Office of High-Energy Physics; the Laboratory Directed Research and Development (LDRD) program of Los Alamos National Laboratory; Consejo Nacional de Ciencia y Tecnolog\'{i}a (CONACyT), M\'{e}xico, grants 271051, 232656, 260378, 179588, 254964, 258865, 243290, 132197, A1-S-46288, A1-S-22784, CF-2023-I-645, c\'{a}tedras 873, 1563, 341, 323, Red HAWC, M\'{e}xico; DGAPA-UNAM grants IG101323, IN111716-3, IN111419, IA102019, IN106521, IN110621, IN110521 , IN102223; VIEP-BUAP; PIFI 2012, 2013, PROFOCIE 2014, 2015; the University of Wisconsin Alumni Research Foundation; the Institute of Geophysics, Planetary Physics, and Signatures at Los Alamos National Laboratory; Polish Science Centre grant, DEC-2017/27/B/ST9/02272; Coordinaci\'{o}n de la Investigaci\'{o}n Cient\'{i}fica de la Universidad Michoacana; Royal Society - Newton Advanced Fellowship 180385; Generalitat Valenciana, grant CIDEGENT/2018/034; The Program Management Unit for Human Resources \& Institutional Development, Research and Innovation, NXPO (grant number B16F630069); Coordinaci\'{o}n General Acad\'{e}mica e Innovaci\'{o}n (CGAI-UdeG), PRODEP-SEP UDG-CA-499; Institute of Cosmic Ray Research (ICRR), University of Tokyo. H.F. acknowledges support by NASA under award number 80GSFC21M0002. We also acknowledge the significant contributions over many years of Stefan Westerhoff, Gaurang Yodh and Arnulfo Zepeda Dominguez, all deceased members of the HAWC collaboration. Thanks to Scott Delay, Luciano D\'{i}az and Eduardo Murrieta for technical support.

\bibliographystyle{aasjournal}

\begin{thebibliography}{}
\expandafter\ifx\csname natexlab\endcsname\relax\def\natexlab#1{#1}\fi
\providecommand{\url}[1]{\href{#1}{#1}}
\providecommand{\dodoi}[1]{doi:~\href{http://doi.org/#1}{\nolinkurl{#1}}}
\providecommand{\doeprint}[1]{\href{http://ascl.net/#1}{\nolinkurl{http://ascl.net/#1}}}
\providecommand{\doarXiv}[1]{\href{https://arxiv.org/abs/#1}{\nolinkurl{https://arxiv.org/abs/#1}}}

\bibitem[{{Abeysekara} {et~al.}(2017){Abeysekara}, {Albert}, {Alfaro}, {Alvarez}, {{\'A}lvarez}, {Arceo}, {Arteaga-Vel{\'a}zquez}, {Ayala Solares}, {Barber}, {Baughman}, {Bautista-Elivar}, {Becerra Gonzalez}, {Becerril}, {Belmont-Moreno}, {BenZvi}, {Berley}, {Bernal}, {Braun}, {Brisbois}, {Caballero-Mora}, {Capistr{\'a}n}, {Carrami{\~n}ana}, {Casanova}, {Castillo}, {Cotti}, {Cotzomi}, {Couti{\~n}o de Le{\'o}n}, {de la Fuente}, {De Le{\'o}n}, {Diaz Hernandez}, {Dingus}, {DuVernois}, {D{\'\i}az-V{\'e}lez}, {Ellsworth}, {Engel}, {Fiorino}, {Fraija}, {Garc{\'\i}a-Gonz{\'a}lez}, {Garfias}, {Gerhardt}, {Gonz{\'a}lez Mu{\~n}oz}, {Gonz{\'a}lez}, {Goodman}, {Hampel-Arias}, {Harding}, {Hernandez}, {Hernandez-Almada}, {Hinton}, {Hui}, {H{\"u}ntemeyer}, {Iriarte}, {Jardin-Blicq}, {Joshi}, {Kaufmann}, {Kieda}, {Lara}, {Lauer}, {Lee}, {Lennarz}, {Le{\'o}n Vargas}, {Linnemann}, {Longinotti}, {Raya}, {Luna-Garc{\'\i}a}, {L{\'o}pez-Coto}, {Malone}, {Marinelli}, {Martinez}, {Martinez-Castellanos}, {Mart{\'\i}nez-Castro},
  {Mart{\'\i}nez-Huerta}, {Matthews}, {Miranda-Romagnoli}, {Moreno}, {Mostaf{\'a}}, {Nellen}, {Newbold}, {Nisa}, {Noriega-Papaqui}, {Pelayo}, {Pretz}, {P{\'e}rez-P{\'e}rez}, {Ren}, {Rho}, {Rivi{\`e}re}, {Rosa-Gonz{\'a}lez}, {Rosenberg}, {Ruiz-Velasco}, {Salazar}, {Salesa Greus}, {Sandoval}, {Schneider}, {Schoorlemmer}, {Sinnis}, {Smith}, {Springer}, {Surajbali}, {Taboada}, {Tibolla}, {Tollefson}, {Torres}, {Ukwatta}, {Vianello}, {Villase{\~n}or}, {Weisgarber}, {Westerhoff}, {Wisher}, {Wood}, {Yapici}, {Younk}, {Zepeda}, \& {Zhou}}]{2HWC}
{Abeysekara}, A.~U., {Albert}, A., {Alfaro}, R., {et~al.} 2017, \apj, 843, 40, \dodoi{10.3847/1538-4357/aa7556}

\bibitem[{Abeysekara {et~al.}(2019)}]{crab2019}
Abeysekara, A.~U., {et~al.} 2019, Astrophys. J., 881, 134, \dodoi{10.3847/1538-4357/ab2f7d}

\bibitem[{{Abeysekara} {et~al.}(2020){Abeysekara}, {Albert}, {Alfaro}, {Angeles Camacho}, {Arteaga-Vel{\'a}zquez}, {Arunbabu}, {Avila Rojas}, {Ayala Solares}, {Baghmanyan}, {Belmont-Moreno}, {BenZvi}, {Brisbois}, {Caballero-Mora}, {Capistr{\'a}n}, {Carrami{\~n}ana}, {Casanova}, {Cotti}, {Cotzomi}, {Couti{\~n}o de Le{\'o}n}, {De la Fuente}, {de Le{\'o}n}, {Dichiara}, {Dingus}, {DuVernois}, {D{\'\i}az-V{\'e}lez}, {Ellsworth}, {Engel}, {Espinoza}, {Fleischhack}, {Fraija}, {Galv{\'a}n-G{\'a}mez}, {Garcia}, {Garc{\'\i}a-Gonz{\'a}lez}, {Garfias}, {Gonz{\'a}lez}, {Goodman}, {Harding}, {Hernandez}, {Hinton}, {Hona}, {Huang}, {Hueyotl-Zahuantitla}, {H{\"u}ntemeyer}, {Iriarte}, {Jardin-Blicq}, {Joshi}, {Kaufmann}, {Kieda}, {Lara}, {Lee}, {Le{\'o}n Vargas}, {Linnemann}, {Longinotti}, {Luis-Raya}, {Lundeen}, {L{\'o}pez-Coto}, {Malone}, {Marinelli}, {Martinez}, {Martinez-Castellanos}, {Mart{\'\i}nez-Castro}, {Mart{\'\i}nez-Huerta}, {Matthews}, {Miranda-Romagnoli}, {Morales-Soto}, {Moreno}, {Mostaf{\'a}}, {Nayerhoda},
  {Nellen}, {Newbold}, {Nisa}, {Noriega-Papaqui}, {Peisker}, {P{\'e}rez-P{\'e}rez}, {Pretz}, {Ren}, {Rho}, {Rivi{\`e}re}, {Rosa-Gonz{\'a}lez}, {Rosenberg}, {Ruiz-Velasco}, {Salesa Greus}, {Sandoval}, {Schneider}, {Schoorlemmer}, {Sinnis}, {Smith}, {Springer}, {Surajbali}, {Tabachnick}, {Tanner}, {Tibolla}, {Tollefson}, {Torres}, {Torres-Escobedo}, {Villase{\~n}or}, {Weisgarber}, {Wood}, {Yapici}, {Zhang}, {Zhou}, \& {HAWC Collaboration}}]{ehwc}
{Abeysekara}, A.~U., {Albert}, A., {Alfaro}, R., {et~al.} 2020, \prl, 124, 021102, \dodoi{10.1103/PhysRevLett.124.021102}

\bibitem[{{Abeysekara} {et~al.}(2022){Abeysekara}, {Albert}, {Alfaro}, {Alvarez}, {{\'A}lvarez Romero}, {Camacho}, {Arteaga Velazquez}, {Kollamparambil}, {Avila Rojas}, {Ayala Solares}, {Babu}, {Baghmanyan}, {Barber}, {Becerra Gonzalez}, {Belmont-Moreno}, {BenZvi}, {Berley}, {Brisbois}, {Caballero Mora}, {Capistr{\'a}n}, {Carrami{\~n}ana}, {Casanova}, {Chaparro-Amaro}, {Cotti}, {Cotzomi}, {Couti{\~n}o de Leon}, {de la Fuente}, {de Le{\'o}n}, {Diaz}, {Diaz Hernandez}, {D{\'\i}az V{\'e}lez}, {Dingus}, {Durocher}, {DuVernois}, {Ellsworth}, {Engel}, {Espinoza Hern{\'a}ndez}, {Fan}, {Fang}, {Fernandez Alonso}, {Fick}, {Fleischhack}, {Flores}, {Fraija}, {Garcia Aguilar}, {Garcia-Gonzalez}, {Garc{\'\i}a-Luna}, {Garc{\'\i}a-Torales}, {Garfias}, {Giacinti}, {Goksu}, {Gonz{\'a}lez}, {Goodman}, {Harding}, {Hern{\'a}ndez Cadena}, {Herzog}, {Hinton}, {Hona}, {Huang}, {Hueyotl-Zahuantitla}, {Hui}, {Humensky}, {H{\"u}ntemeyer}, {Iriarte}, {Jardin-Blicq}, {Jhee}, {Joshi}, {Kieda}, {Kunde}, {Kunwar}, {Lara}, {Lee}, {Lee},
  {Lennarz}, {Vargas}, {Linnemann}, {Longinotti}, {Lopez-Coto}, {Luis-Raya}, {Lundeen}, {Malone}, {Marandon}, {Martinez}, {Martinez Castellanos}, {Mart{\'\i}nez Huerta}, {Mart{\'\i}nez-Castro}, {Matthews}, {McEnery}, {Miranda-Romagnoli}, {Morales Soto}, {Moreno Barbosa}, {Mostafa}, {Nayerhoda}, {Nellen}, {Newbold}, {Nisa}, {Noriega-Papaqui}, {Olivera-Nieto}, {Omodei}, {Peisker}, {P{\'e}rez Araujo}, {P{\'e}rez P{\'e}rez}, {Rho}, {Rivi{\`e}re}, {Rosa-Gonzalez}, {Ruiz-Velasco}, {Ryan}, {Salazar}, {Salesa Greus}, {Sandoval}, {Schneider}, {Schoorlemmer}, {Serna-Franco}, {Sinnis}, {Smith}, {Springer}, {Surajbali}, {Taboada}, {Tanner}, {Tollefson}, {Torres}, {Torres Escobedo}, {Turner}, {Ure{\~n}a-Mena}, {Villase{\~n}or}, {Wang}, {Watson}, {Weisgarber}, {Werner}, {Willox}, {Wood}, {Yodh}, {Zepeda}, \& {Zhou}}]{chadicrc}
{Abeysekara}, A.~U., {Albert}, A., {Alfaro}, R., {et~al.} 2022, in 37th International Cosmic Ray Conference, 828, \dodoi{10.22323/1.395.0828}

\bibitem[{{Abeysekara} {et~al.}(2023){Abeysekara}, {Albert}, {Alfaro}, {Alvarez}, {{\'A}lvarez}, {Araya}, {Arteaga-Vel{\'a}zquez}, {Arunbabu}, {Avila Rojas}, {Ayala Solares}, {Babu}, {Barber}, {Becerril}, {Belmont-Moreno}, {BenZvi}, {Blanco}, {Braun}, {Brisbois}, {Caballero-Mora}, {Cabrera Mart{\'\i}nez}, {Capistr{\'a}n}, {Carrami{\~n}ana}, {Casanova}, {Castillo}, {Chaparro-Amaro}, {Cotti}, {Cotzomi}, {Couti{\~n}o de Le{\'o}n}, {de la Fuente}, {de Le{\'o}n}, {De Young}, {Hernandez}, {Dingus}, {DuVernois}, {Durocher}, {D{\'\i}az-V{\'e}lez}, {Ellsworth}, {Engel}, {Espinoza}, {Fan}, {Fang}, {Fick}, {Fleischhack}, {Flores}, {Fraija}, {Garc{\'\i}a-Gonz{\'a}lez}, {Garcia-Torales}, {Garfias}, {Giacinti}, {Goksu}, {Gonz{\'a}lez}, {Gonz{\'a}lez-Mu{\~n}oz}, {Goodman}, {Harding}, {Hernandez}, {Hernandez}, {Hinton}, {Hona}, {Huang}, {Hueyotl-Zahuantitla}, {Hui}, {Humensky}, {H{\"u}ntemeyer}, {Iriarte}, {Imran}, {Jardin-Blicq}, {Joshi}, {Kaufmann}, {Kieda}, {Kunde}, {Lara}, {Lauer}, {Lee}, {Lennarz}, {Vargas},
  {Linnemann}, {Longinotti}, {Luis-Raya}, {Lundeen}, {Malone}, {Marandon}, {Marinelli}, {Martinez}, {Mart{\'\i}nez-Castellanos}, {Mart{\'\i}nez-Castro}, {Mart{\'\i}nez-Huerta}, {Matthews}, {Miranda-Romagnoli}, {Montaruli}, {Morales-Soto}, {Moreno}, {Mostaf{\'a}}, {Nayerhoda}, {Nellen}, {Newbold}, {Nisa}, {Noriega-Papaqui}, {Oceguera-Becerra}, {Olivera-Nieto}, {Omodei}, {Peisker}, {P{\'e}rez Araujo}, {P{\'e}rez-P{\'e}rez}, {Ponce}, {Pretz}, {Rho}, {Rosa-Gonz{\'a}lez}, {Ruiz-Velasco}, {Salazar}, {Salazar-Gallegos}, {Salesa Greus}, {Sandoval}, {Schneider}, {Schoorlemmer}, {Serna-Franco}, {Sinnis}, {Smith}, {Son}, {Sparks Woodle}, {Springer}, {Taboada}, {Tepe}, {Tibolla}, {Tollefson}, {Torres}, {Torres-Escobedo}, {Turner}, {Ure{\~n}a-Mena}, {Ukwatta}, {Varela}, {Vargas-Maga{\~n}a}, {Villase{\~n}or}, {Wang}, {Watson}, {Werner}, {Westerhoff}, {Willox}, {Wisher}, {Wood}, {Yodh}, {Zaborov}, {Zepeda}, {Zhou}, \& {HAWC Collaboration}}]{nim}
---. 2023, Nuclear Instruments and Methods in Physics Research A, 1052, 168253, \dodoi{10.1016/j.nima.2023.168253}

\bibitem[{{Ackermann} {et~al.}(2017){Ackermann}, {Ajello}, {Baldini}, {Ballet}, {Barbiellini}, {Bastieri}, {Bellazzini}, {Bissaldi}, {Bloom}, {Bonino}, {Bottacini}, {Brandt}, {Bregeon}, {Bruel}, {Buehler}, {Cameron}, {Caragiulo}, {Caraveo}, {Castro}, {Cavazzuti}, {Cecchi}, {Charles}, {Chekhtman}, {Cheung}, {Chiaro}, {Ciprini}, {Cohen}, {Costantin}, {Costanza}, {Cutini}, {D'Ammando}, {de Palma}, {Desiante}, {Digel}, {Di Lalla}, {Di Mauro}, {Di Venere}, {Favuzzi}, {Fegan}, {Ferrara}, {Franckowiak}, {Fukazawa}, {Funk}, {Fusco}, {Gargano}, {Gasparrini}, {Giglietto}, {Giordano}, {Giroletti}, {Green}, {Grenier}, {Grondin}, {Guillemot}, {Guiriec}, {Harding}, {Hays}, {Hewitt}, {Horan}, {Hou}, {J{\'o}hannesson}, {Kamae}, {Kuss}, {La Mura}, {Larsson}, {Lemoine-Goumard}, {Li}, {Longo}, {Loparco}, {Lubrano}, {Magill}, {Maldera}, {Malyshev}, {Manfreda}, {Mazziotta}, {Michelson}, {Mitthumsiri}, {Mizuno}, {Monzani}, {Morselli}, {Moskalenko}, {Negro}, {Nuss}, {Ohsugi}, {Omodei}, {Orienti}, {Orlando}, {Ormes}, {Paliya},
  {Paneque}, {Perkins}, {Persic}, {Pesce-Rollins}, {Petrosian}, {Piron}, {Porter}, {Principe}, {Rain{\`o}}, {Rando}, {Razzano}, {Razzaque}, {Reimer}, {Reimer}, {Reposeur}, {Sgr{\`o}}, {Simone}, {Siskind}, {Spada}, {Spandre}, {Spinelli}, {Suson}, {Tak}, {Thayer}, {Thompson}, {Torres}, {Tosti}, {Troja}, {Vianello}, {Wood}, \& {Wood}}]{4FGES}
{Ackermann}, M., {Ajello}, M., {Baldini}, L., {et~al.} 2017, \apj, 843, 139, \dodoi{10.3847/1538-4357/aa775a}

\bibitem[{{Aharonian} {et~al.}(2007){Aharonian}, {Akhperjanian}, {Bazer-Bachi}, {Behera}, {Beilicke}, {Benbow}, {Berge}, {Bernl{\"o}hr}, {Boisson}, {Bolz}, {Borrel}, {Braun}, {Brion}, {Brown}, {B{\"u}hler}, {B{\"u}sching}, {Boutelier}, {Carrigan}, {Chadwick}, {Chounet}, {Coignet}, {Cornils}, {Costamante}, {Degrange}, {Dickinson}, {Djannati-Ata{\"\i}}, {Domainko}, {Drury}, {Dubus}, {Egberts}, {Emmanoulopoulos}, {Espigat}, {Farnier}, {Feinstein}, {Fiasson}, {F{\"o}rster}, {Fontaine}, {Funk}, {Funk}, {F{\"u}{\ss}ling}, {Gallant}, {Giebels}, {Glicenstein}, {Gl{\"u}ck}, {Goret}, {Hadjichristidis}, {Hauser}, {Hauser}, {Heinzelmann}, {Henri}, {Hermann}, {Hinton}, {Hoffmann}, {Hofmann}, {Holleran}, {Hoppe}, {Horns}, {Jacholkowska}, {de Jager}, {Kendziorra}, {Kerschhaggl}, {Kh{\'e}lifi}, {Komin}, {Kosack}, {Lamanna}, {Latham}, {Le Gallou}, {Lemi{\`e}re}, {Lemoine-Goumard}, {Lohse}, {Martin}, {Martineau-Huynh}, {Marcowith}, {Masterson}, {Maurin}, {McComb}, {Moulin}, {de Naurois}, {Nedbal}, {Nolan}, {Noutsos}, {Olive},
  {Orford}, {Osborne}, {Panter}, {Pedaletti}, {Pelletier}, {Petrucci}, {Pita}, {P{\"u}hlhofer}, {Punch}, {Ranchon}, {Raubenheimer}, {Raue}, {Rayner}, {Ripken}, {Rob}, {Rolland}, {Rosier-Lees}, {Rowell}, {Ruppel}, {Sahakian}, {Santangelo}, {Saug{\'e}}, {Schlenker}, {Schlickeiser}, {Schr{\"o}der}, {Schwanke}, {Schwarzburg}, {Schwemmer}, {Shalchi}, {Sol}, {Spangler}, {Steenkamp}, {Stegmann}, {Superina}, {Tam}, {Tavernet}, {Terrier}, {Tluczykont}, {van Eldik}, {Vasileiadis}, {Venter}, {Vialle}, {Vincent}, {V{\"o}lk}, {Wagner}, \& {Ward}}]{Originpaper}
{Aharonian}, F., {Akhperjanian}, A.~G., {Bazer-Bachi}, A.~R., {et~al.} 2007, \aap, 472, 489, \dodoi{10.1051/0004-6361:20077280}

\bibitem[{Albert {et~al.}(2020)}]{3HWC}
Albert, A., {et~al.} 2020, Astrophys. J., 905, 76, \dodoi{10.3847/1538-4357/abc2d8}

\bibitem[{{Albert} {et~al.}(2021){Albert}, {Alfaro}, {Alvarez}, {Arteaga-Vel{\`a}zquez}, {Arunbabu}, {Avila Rojas}, {Ayala Solares}, {Baghmanyan}, {Belmont-Moreno}, {Brisbois}, {Caballero-Mora}, {Capistr{\`a}n}, {Carrami{\~n}ana}, {Casanova}, {Cotzomi}, {Coutin{\~o} de Le{\'o}n}, {De la Fuente}, {Diaz Hernandez}, {Dingus}, {DuVernois}, {Durocher}, {Engel}, {Espinoza}, {Fraija}, {Garcia}, {Garc{\'\i}a-Gonz{\'a}lez}, {Giacinti}, {Gonz{\'a}lez}, {Goodman}, {Harding}, {Hinton}, {Hona}, {Huang}, {Hueyotl-Zahuantitla}, {Huentemeyer}, {Jardin-Blicq}, {Joshi}, {Lee}, {Le{\'o}n Vargas}, {Linnemann}, {Longinotti}, {Luis-Raya}, {Lundeen}, {L{\'o}pez-Coto}, {Malone}, {Martinez}, {Mart{\'\i}nez-Castro}, {Matthews}, {Miranda-Romagnoli}, {Morales-Soto}, {Moreno}, {Mostaf{\'a}}, {Nayerhoda}, {Nellen}, {Newbold}, {Nisa}, {Noriega-Papaqui}, {Olivera-Nieto}, {Omodei}, {Peisker}, {P{\'e}rez Araujo}, {P{\'e}rez-P{\'e}rez}, {Rho}, {Rosa-Gonz{\`a}lez}, {Ruiz-Velasco}, {Salazar}, {Salesa Greus}, {Sandoval}, {Schneider},
  {Schoorlemmer}, {Serna-Franco}, {Smith}, {Springer}, {Surajbali}, {Tollefson}, {Torres}, {Turner}, {Uren{\~a}-Mena}, {Weisgarber}, {Willox}, {Zhou}, \& {de Le{\'o}n}}]{j2019}
{Albert}, A., {Alfaro}, R., {Alvarez}, C., {et~al.} 2021, arXiv e-prints, arXiv:2101.01649, \dodoi{10.48550/arXiv.2101.01649}

\bibitem[{{Anada} {et~al.}(2010){Anada}, {Bamba}, {Ebisawa}, \& {Dotani}}]{xraysuzaku}
{Anada}, T., {Bamba}, A., {Ebisawa}, K., \& {Dotani}, T. 2010, \pasj, 62, 179, \dodoi{10.1093/pasj/62.1.179}

\bibitem[{{Araya}(2018)}]{araya}
{Araya}, M. 2018, \apj, 859, 69, \dodoi{10.3847/1538-4357/aabd7e}

\bibitem[{Ballet {et~al.}(2020)Ballet, Burnett, Digel, \& Lott}]{4fgl2}
Ballet, J., Burnett, T.~H., Digel, S.~W., \& Lott, B. 2020, Fermi Large Area Telescope Fourth Source Catalog Data Release 2,  arXiv, \dodoi{10.48550/ARXIV.2005.11208}

\bibitem[{{Bamba} {et~al.}(2003){Bamba}, {Ueno}, {Koyama}, \& {Yamauchi}}]{xraysnr}
{Bamba}, A., {Ueno}, M., {Koyama}, K., \& {Yamauchi}, S. 2003, \apj, 589, 253, \dodoi{10.1086/374354}

\bibitem[{{Boxi} \& {Gupta}(2024)}]{boxi}
{Boxi}, S., \& {Gupta}, N. 2024, \apj, 961, 61, \dodoi{10.3847/1538-4357/ad0da9}

\bibitem[{Brogan {et~al.}(2004)Brogan, Devine, Lazio, Kassim, Tam, Brisken, Dyer, \& Roberts}]{snrradio}
Brogan, C.~L., Devine, K.~E., Lazio, T.~J., {et~al.} 2004, Astron. J., 127, 355, \dodoi{10.1086/379856}

\bibitem[{{Brogan} {et~al.}(2004){Brogan}, {Devine}, {Lazio}, {Kassim}, {Tam}, {Brisken}, {Dyer}, \& {Roberts}}]{snrg111}
{Brogan}, C.~L., {Devine}, K.~E., {Lazio}, T.~J., {et~al.} 2004, \aj, 127, 355, \dodoi{10.1086/379856}

\bibitem[{{Brogan} {et~al.}(2006){Brogan}, {Gelfand}, {Gaensler}, {Kassim}, \& {Lazio}}]{brogan}
{Brogan}, C.~L., {Gelfand}, J.~D., {Gaensler}, B.~M., {Kassim}, N.~E., \& {Lazio}, T.~J.~W. 2006, \apjl, 639, L25, \dodoi{10.1086/501500}

\bibitem[{{Cao} {et~al.}(2023){Cao}, {Aharonian}, {An}, {Axikegu}, {Bai}, {Bao}, {Bastieri}, {Bi}, {Bi}, {Cai}, {Cao}, {Cao}, {Cao}, {Chang}, {Chang}, {Chen}, {Chen}, {Chen}, {Chen}, {Chen}, {Chen}, {Chen}, {Chen}, {Chen}, {Chen}, {Chen}, {Chen}, {Cheng}, {Cheng}, {Cui}, {Cui}, {Cui}, {Cui}, {Dai}, {Dai}, {Dai}, {Danzengluobu}, {della Volpe}, {Dong}, {Duan}, {Fan}, {Fan}, {Fang}, {Fang}, {Feng}, {Feng}, {Feng}, {Feng}, {Feng}, {Gabici}, {Gao}, {Gao}, {Gao}, {Gao}, {Gao}, {Gao}, {Ge}, {Geng}, {Giacinti}, {Gong}, {Gou}, {Gu}, {Guo}, {Guo}, {Guo}, {Guo}, {Han}, {He}, {He}, {He}, {He}, {He}, {Heller}, {Hor}, {Hou}, {Hou}, {Hou}, {Hu}, {Hu}, {Hu}, {Huang}, {Huang}, {Huang}, {Huang}, {Huang}, {Huang}, {Huang}, {Ji}, {Jia}, {Jia}, {Jiang}, {Jiang}, {Jiang}, {Jin}, {Kang}, {Ke}, {Kuleshov}, {Kurinov}, {Li}, {Li}, {Li}, {Li}, {Li}, {Li}, {Li}, {Li}, {Li}, {Li}, {Li}, {Li}, {Li}, {Li}, {Li}, {Li}, {Li}, {Li}, {Li}, {Liang}, {Liang}, {Lin}, {Liu}, {Liu}, {Liu}, {Liu}, {Liu}, {Liu}, {Liu}, {Liu}, {Liu}, {Liu}, {Liu},
  {Liu}, {Liu}, {Liu}, {Lu}, {Luo}, {Lv}, {Ma}, {Ma}, {Ma}, {Mao}, {Min}, {Mitthumsiri}, {Mu}, {Nan}, {Neronov}, {Ou}, {Pang}, {Pattarakijwanich}, {Pei}, {Qi}, {Qi}, {Qiao}, {Qin}, {Ruffolo}, {S{\'a}iz}, {Semikoz}, {Shao}, {Shao}, {Shchegolev}, {Sheng}, {Shu}, {Song}, {Stenkin}, {Stepanov}, {Su}, {Sun}, {Sun}, {Sun}, {Tam}, {Tang}, {Tang}, {Tian}, {Wang}, {Wang}, {Wang}, {Wang}, {Wang}, {Wang}, {Wang}, {Wang}, {Wang}, {Wang}, {Wang}, {Wang}, {Wang}, {Wang}, {Wang}, {Wang}, {Wang}, {Wang}, {Wang}, {Wang}, {Wang}, {Wei}, {Wei}, {Wei}, {Wen}, {Wu}, {Wu}, {Wu}, {Wu}, {Wu}, {Xi}, {Xia}, {Xia}, {Xiang}, {Xiao}, {Xiao}, {Xin}, {Xin}, {Xing}, {Xiong}, {Xu}, {Xu}, {Xu}, {Xu}, {Xue}, {Yan}, {Yan}, {Yan}, {Yang}, {Yang}, {Yang}, {Yang}, {Yang}, {Yang}, {Yang}, {Yang}, {Yang}, {Yao}, {Yao}, {Ye}, {Yin}, {Yin}, {You}, {You}, {Yu}, {Yuan}, {Yue}, {Zeng}, {Zeng}, {Zeng}, {Zha}, {Zhang}, {Zhang}, {Zhang}, {Zhang}, {Zhang}, {Zhang}, {Zhang}, {Zhang}, {Zhang}, {Zhang}, {Zhang}, {Zhang}, {Zhang}, {Zhang}, {Zhang}, {Zhang},
  {Zhang}, {Zhang}, {Zhao}, {Zhao}, {Zhao}, {Zhao}, {Zhao}, {Zheng}, {Zhou}, {Zhou}, {Zhou}, {Zhou}, {Zhou}, {Zhou}, {Zhou}, {Zhu}, {Zhu}, {Zhu}, {Zhu}, \& {Zuo.}}]{lhaasocat}
{Cao}, Z., {Aharonian}, F., {An}, Q., {et~al.} 2023, arXiv e-prints, arXiv:2305.17030, \dodoi{10.48550/arXiv.2305.17030}

\bibitem[{{Castelletti, G.} {et~al.}(2016){Castelletti, G.}, {Giacani, E.}, \& {Petriella, A.}}]{castelleti16}
{Castelletti, G.}, {Giacani, E.}, \& {Petriella, A.} 2016, A\&A, 587, A71, \dodoi{10.1051/0004-6361/201527578}

\bibitem[{{Dermer} \& {Powale}(2013)}]{snreff}
{Dermer}, C.~D., \& {Powale}, G. 2013, \aap, 553, A34, \dodoi{10.1051/0004-6361/201220394}

\bibitem[{{Gaensler} \& {Slane}(2006)}]{gamerapwnevol}
{Gaensler}, B.~M., \& {Slane}, P.~O. 2006, \araa, 44, 17, \dodoi{10.1146/annurev.astro.44.051905.092528}

\bibitem[{Ginzburg \& Ptuskin(1976)}]{snrenergy}
Ginzburg, V.~L., \& Ptuskin, V.~S. 1976, Rev. Mod. Phys., 48, 161, \dodoi{10.1103/RevModPhys.48.161}

\bibitem[{{Hahn} {et~al.}(2022){Hahn}, {Romoli}, \& {Breuhaus}}]{gamera}
{Hahn}, J., {Romoli}, C., \& {Breuhaus}, M. 2022, {GAMERA: Source modeling in gamma astronomy}, Astrophysics Source Code Library, record ascl:2203.007.
\newblock \doeprint{2203.007}

\bibitem[{{H.E.S.S. Collaboration} {et~al.}(2023){H.E.S.S. Collaboration}, {Aharonian}, {Ait Benkhali}, {Aschersleben}, {Ashkar}, {Backes}, {Barbosa Martins}, {Batzofin}, {Becherini}, {Berge}, {B{\"o}ttcher}, {Boisson}, {Bolmont}, {Borowska}, {Bouyahiaoui}, {Bradascio}, {Breuhaus}, {Brose}, {Brun}, {Bruno}, {Bulik}, {Burger-Scheidlin}, {Bylund}, {Caroff}, {Casanova}, {Celic}, {Cerruti}, {Chambery}, {Chand}, {Chen}, {Chibueze}, {Chibueze}, {Damascene Mbarubucyeye}, {Djannati-Ata{\"\i}}, {Dmytriiev}, {Einecke}, {Ernenwein}, {Feijen}, {Filipovic}, {Fontaine}, {F{\"u}{\ss}ling}, {Funk}, {Gabici}, {Gallant}, {Ghafourizadeh}, {Giavitto}, {Giunti}, {Glawion}, {Goswami}, {Grolleron}, {Grondin}, {Haerer}, {Hinton}, {Hofmann}, {Holch}, {Holler}, {Horns}, {Huang}, {Jamrozy}, {Jankowsky}, {Joshi}, {Jung-Richardt}, {Kasai}, {Katarzy{\'n}ski}, {Kh{\'e}lifi}, {Klu{\'z}niak}, {Komin}, {Kosack}, {Kostunin}, {Lang}, {Le Stum}, {Leitl}, {Lemi{\`e}re}, {Lemoine-Goumard}, {Lenain}, {Leuschner}, {Lohse}, {Luashvili}, {Lypova},
  {Mackey}, {Malyshev}, {Malyshev}, {Marandon}, {Marchegiani}, {Marcowith}, {Marinos}, {Mart{\'\i}-Devesa}, {Marx}, {Mitchell}, {Moderski}, {Mohrmann}, {Montanari}, {Moulin}, {Muller}, {Nakashima}, {de Naurois}, {Niemiec}, {Priyana Noel}, {Ohm}, {Olivera-Nieto}, {de Ona Wilhelmi}, {Ostrowski}, {Panny}, {Panter}, {Parsons}, {Prokhorov}, {P{\"u}hlhofer}, {Punch}, {Quirrenbach}, {Reichherzer}, {Reimer}, {Reimer}, {Renaud}, {Reville}, {Rieger}, {Rowell}, {Rudak}, {Sahakian}, {Santangelo}, {Sasaki}, {Schutte}, {Schwanke}, {Shapopi}, {Sol}, {Specovius}, {Spencer}, {Stawarz}, {Steenkamp}, {Steinmassl}, {Sushch}, {Suzuki}, {Takahashi}, {Tanaka}, {Terrier}, {Thorpe-Morgan}, {Tsirou}, {Tsuji}, {Uchiyama}, {van Eldik}, {Vecchi}, {Veh}, {Venter}, {Vink}, {Wach}, {Wagner}, {White}, {Wierzcholska}, {Wong}, {Zacharias}, {Zargaryan}, {Zdziarski}, {Zech}, {Zouari}, \& {{\.Z}ywucka}}]{new1809hesspaper}
{H.E.S.S. Collaboration}, {Aharonian}, F., {Ait Benkhali}, F., {et~al.} 2023, \aap, 672, A103, \dodoi{10.1051/0004-6361/202245459}

\bibitem[{{Manchester} {et~al.}(2005){Manchester}, {Hobbs}, {Teoh}, \& {Hobbs}}]{atnf}
{Manchester}, R.~N., {Hobbs}, G.~B., {Teoh}, A., \& {Hobbs}, M. 2005, \aj, 129, 1993, \dodoi{10.1086/428488}

\bibitem[{{Marandon} {et~al.}(2019){Marandon}, {Jardin-Blicq}, \& {Schoorlemmer}}]{outriggericrc}
{Marandon}, V., {Jardin-Blicq}, A., \& {Schoorlemmer}, H. 2019, in International Cosmic Ray Conference, Vol.~36, 36th International Cosmic Ray Conference (ICRC2019), 736, \dodoi{10.22323/1.358.0736}

\bibitem[{{Morris} {et~al.}(2002){Morris}, {Hobbs}, {Lyne}, {Stairs}, {Camilo}, {Manchester}, {Possenti}, {Bell}, {Kaspi}, {Amico}, {McKay}, {Crawford}, \& {Kramer}}]{parkespulsar}
{Morris}, D.~J., {Hobbs}, G., {Lyne}, A.~G., {et~al.} 2002, \mnras, 335, 275, \dodoi{10.1046/j.1365-8711.2002.05551.x}

\bibitem[{{Popescu} {et~al.}(2017){Popescu}, {Yang}, {Tuffs}, {Natale}, {Rushton}, \& {Aharonian}}]{radfield}
{Popescu}, C.~C., {Yang}, R., {Tuffs}, R.~J., {et~al.} 2017, \mnras, 470, 2539, \dodoi{10.1093/mnras/stx1282}

\bibitem[{{Rangelov} {et~al.}(2014){Rangelov}, {Posselt}, {Kargaltsev}, {Pavlov}, {Hare}, \& {Volkov}}]{psrj1811}
{Rangelov}, B., {Posselt}, B., {Kargaltsev}, O., {et~al.} 2014, \apj, 796, 34, \dodoi{10.1088/0004-637X/796/1/34}

\bibitem[{{Shan} {et~al.}(2018){Shan}, {Zhu}, {Tian}, {Zhang}, {Zhang}, {Wu}, \& {Yang}}]{shansnr}
{Shan}, S.~S., {Zhu}, H., {Tian}, W.~W., {et~al.} 2018, \apjs, 238, 35, \dodoi{10.3847/1538-4365/aae07a}

\bibitem[{{Torres} {et~al.}(2003){Torres}, {Romero}, {Dame}, {Combi}, \& {Butt}}]{gammaeqn}
{Torres}, D.~F., {Romero}, G.~E., {Dame}, T.~M., {Combi}, J.~A., \& {Butt}, Y.~M. 2003, \physrep, 382, 303, \dodoi{10.1016/S0370-1573(03)00201-1}

\bibitem[{{Umemoto} {et~al.}(2017){Umemoto}, {Minamidani}, {Kuno}, {Fujita}, {Matsuo}, {Nishimura}, {Torii}, {Tosaki}, {Kohno}, {Kuriki}, {Tsuda}, {Hirota}, {Ohashi}, {Yamagishi}, {Handa}, {Nakanishi}, {Omodaka}, {Koide}, {Matsumoto}, {Onishi}, {Tokuda}, {Seta}, {Kobayashi}, {Tachihara}, {Sano}, {Hattori}, {Onodera}, {Oasa}, {Kamegai}, {Tsuboi}, {Sofue}, {Higuchi}, {Chibueze}, {Mizuno}, {Honma}, {Muller}, {Inoue}, {Morokuma-Matsui}, {Shinnaga}, {Ozawa}, {Takahashi}, {Yoshiike}, {Costes}, \& {Kuwahara}}]{fugin}
{Umemoto}, T., {Minamidani}, T., {Kuno}, N., {et~al.} 2017, \pasj, 69, 78, \dodoi{10.1093/pasj/psx061}

\bibitem[{{Vianello} {et~al.}(2015){Vianello}, {Lauer}, {Younk}, {Tibaldo}, {Burgess}, {Ayala}, {Harding}, {Hui}, {Omodei}, \& {Zhou}}]{Vianello}
{Vianello}, G., {Lauer}, R.~J., {Younk}, P., {et~al.} 2015, arXiv e-prints, arXiv:1507.08343, \dodoi{10.48550/arXiv.1507.08343}

\bibitem[{{Voisin} {et~al.}(2019){Voisin}, {Rowell}, {Burton}, {Fukui}, {Sano}, {Aharonian}, {Maxted}, {Braiding}, {Blackwell}, \& {Lau}}]{viosin}
{Voisin}, F.~J., {Rowell}, G.~P., {Burton}, M.~G., {et~al.} 2019, \pasa, 36, e014, \dodoi{10.1017/pasa.2019.7}

\bibitem[{Wilks(1938)}]{wilks}
Wilks, S.~S. 1938, The Annals of Mathematical Statistics, 9, 60 , \dodoi{10.1214/aoms/1177732360}

\bibitem[{{Zabalza}(2015)}]{naima}
{Zabalza}, V. 2015, in International Cosmic Ray Conference, Vol.~34, 34th International Cosmic Ray Conference (ICRC2015), 922, \dodoi{10.22323/1.236.0922}

\end{thebibliography}

\end{document}